\documentclass[reprint,superscriptaddress,amsmath,amssymb,aps,prl,longbibliography]{revtex4-1}
\usepackage{hyperref}
\usepackage{graphicx}
\usepackage{bm}
\usepackage{bbold}

\begin{document}
\title{Pulsed Quantum-State Reconstruction of Dark Systems}
\author{Yu Liu}
\author{Jiazhao Tian}
\affiliation{School of Physics, Huazhong University of Science and Technology, Wuhan 430074, China}
\affiliation{International Joint Laboratory on Quantum Sensing and Quantum Metrology,
Huazhong University of Science and Technology, Wuhan, 430074, China}
\author{Ralf Betzholz}
\email{ralf\_betzholz@hust.edu.cn}
\affiliation{School of Physics, Huazhong University of Science and Technology, Wuhan 430074, China}
\affiliation{International Joint Laboratory on Quantum Sensing and Quantum Metrology,
Huazhong University of Science and Technology, Wuhan, 430074, China}
\author{Jianming Cai}
\email{jianmingcai@hust.edu.cn}
\affiliation{School of Physics, Huazhong University of Science and Technology, Wuhan 430074, China}
\affiliation{International Joint Laboratory on Quantum Sensing and Quantum Metrology,
Huazhong University of Science and Technology, Wuhan, 430074, China}
\date{\today}

\begin{abstract}
We propose a novel strategy to reconstruct the quantum state of dark systems, i.e., degrees of freedom that are not directly accessible for measurement or control. Our scheme relies on the quantum control of a two-level probe that exerts a state-dependent potential on the dark system. Using a sequence of control pulses applied to the probe makes it possible to tailor the information one can obtain and, for example, allows us to reconstruct the density operator of a dark spin as well as the Wigner characteristic function of a harmonic oscillator. Because of the symmetry of the applied pulse sequence, this scheme is robust against slow noise on the probe. The proof-of-principle experiments are readily feasible in solid-state spins and trapped ions.
\end{abstract}
\maketitle

\textit{Introduction.---}The measurement of the quantum state of a system is a prerequisite ingredient in most modern quantum experiments, ranging from fundamental tests of quantum mechanics~\cite{Aspelmeyer2014,Neukirch2015} to various quantum-information-processing tasks~\cite{Nielsen2003,Briegel2009,Ladd2010}. However, even with the rapid progress in the coherent manipulation and quantum-state tomography of several quantum systems, such as photons~\cite{Altepeter2005,Deglise2008}, electron spins~\cite{Elzerman2004,Barthel2009,Morello2010}, atomic qubits~\cite{Bochmann2010}, superconducting circuits~\cite{Steffen2006,Fedorov2010}, and mechanical resonators~\cite{Vanner2013,Rashid2017}, many quantum systems still remain difficult to access for a direct observation of their state, systems we will refer to as dark. In order to circumvent the requirement of such a direct access, a promising technique is to employ an auxiliary quantum system as a measurement probe, on which measurements as well as coherent manipulations can be performed~\cite{Leibfried1996,Bertet2002,Wallraff2005,Burgarth2011,Burgarth2015,Sone2017a,Sone2017b,Zhang2018}. Interferometry~\cite{Cronin2009} based on such a measurement probe allows us to extract information on a target system~\cite{Recati2005,Quan2006,Dorner2013,Peng2015,Correa2015,Asadian2014}. Nevertheless, it still remains a key challenge to achieve a full quantum-state tomography of dark systems without requiring any direct control.

In this Letter, we propose a general scheme for a probe-measurement based quantum-state reconstruction of dark systems, where the obtainable dark-system quantities can be tailored by a pulsed control of the two-level probe we employ. The scheme does not require any manipulation of the dark systems or the controllability of the coupling to the probe. This is a requirement on which, for example, many previous reconstruction methods for continuous-variable systems depend~\cite{Lutterbach1997,Kim1998,Singh2010,Casanova2012,Mazzola2013,Taketani2014}. Additionally, it inherits the feature of robustness against noise on the probe from pulsed dynamical decoupling~\cite{Uhri2007,de-Lange2010,Naydenov2011}, making it suitable also for noisy environments such as solid-state platforms. The proposed strategy is exemplified at the quantum-state tomography of a dark spin-1/2 and a dark harmonic oscillator by reconstructing their density operator and Wigner characteristic function, respectively. We discuss the feasibility of the proof-of-principle experiments to testify the distinct advantages of the present proposal in solid-state spin~\cite{Taminiau2012,Zhao2011,Kolkowitz2012,Liu2017} and trapped-ion systems~\cite{Johanning2009,Khromova2012,Sriarunothai2017} within state-of-the-art experimental capabilities. The scheme is applicable to other dark systems as in a variety of physical settings, making it a versatile tool for quantum measurements. 

\textit{Pulsed state-reconstruction scheme.---}The probe we consider is a generic two-level system described by the Hamiltonian $H_p=(\omega_p/2)\sigma_z^p$, with the Pauli operator $\sigma_z^p=|1\rangle_p\langle1|-|0\rangle_p\langle0|$. We denote the Hamiltonian of the dark system by $H_d$. The underlying idea of the proposed scheme is that the interaction between the probe and the dark system is given by probe-state-dependent potentials $H_0$ and $H_1$ for the dark system, i.e., an interaction of the form $H_{\rm int}=|0\rangle_p\langle 0| H_0+|1\rangle_p\langle 1| H_1$. The dynamics of the combined system is then governed by $H_p+H_d+H_{\rm int}$ and in the interaction picture with respect to $H_p$ this Hamiltonian has the form
\begin{equation}
\label{eq:H_int}
H=|0\rangle_p\langle 0| V_0+|1\rangle_p\langle 1| V_1,
\end{equation}
with $V_0=H_d+H_0$ and $V_1=H_d+H_1$ acting on the dark system. The dynamics generated by this Hamiltonian is used to obtain information about the state of the dark system.

In order to do so, the probe is initialized in the superposition state $|+\rangle_p=(|0\rangle_p+|1\rangle_p)/\sqrt{2}$, such that the initial state of the full system has the form $|\Psi(0)\rangle=|+\rangle_p|\psi\rangle$, with the dark-system state $|\psi\rangle$. The free evolution of this state under the Hamiltonian~\eqref{eq:H_int} generates entanglement between the probe and the dark system and thereby allows to connect measurements on the probe with quantities of the dark system. However, as we will show, appreciably more information can be obtained by the application of a series of pulses that manipulate the probe~\cite{Li2011,Zhao2014}. Explicitly, we apply a series of $2N$  $\pi$ pulses, all separated by the free-evolution time $\tau$, and thereby modulate the effective potential acting on the dark system. After such an evolution of total time $t=2N\tau$ the state of the full system has evolved into
\begin{eqnarray}
\label{eq:Psi_T}
|\Psi(t)\rangle=\frac{1}{\sqrt{2}}\big(|0\rangle_p U_0|\psi\rangle+|1\rangle_p U_1|\psi\rangle\big),
\end{eqnarray}
where the state-dependent dark-system time-evolution operators are given by $U_0=u_0^N$ and $U_1=u_1^N$, with the single pulse-segment evolution operators
\begin{equation}
\begin{aligned}
u_0=& \exp(-iV_1\tau) \exp(-iV_0\tau),\\
u_1=& \exp(-iV_0\tau) \exp(-iV_1\tau).
\end{aligned}
\end{equation}
In order to obtain any information on the dark-system state from this dynamics, a necessary condition is that the operators $V_0$ and $V_1$ do not commute, since otherwise the above evolution operators coincide. Following this time evolution, a measurement of the probe Pauli vector $\boldsymbol{\sigma}_p=(\sigma_x^p,\sigma_y^p,\sigma_z^p)$ can be performed resulting in $\langle \boldsymbol{\sigma}_p\rangle={\rm Tr}\{\boldsymbol{\sigma}_p\varrho(t)\}$,  with $\varrho(t)=|\Psi(t)\rangle\langle\Psi(t)|$. The generalization to initially separable density operators of the form $\varrho(0)=|+\rangle_p\langle+|\,\rho$, with the possibly mixed initial dark-system density operator $\rho$, is straightforward and yields
\begin{equation}
\begin{aligned}
\label{eq:sigma_p}
\langle\sigma_x^p\rangle=&\frac{1}{2}{\rm Tr}\big\{\big(U_0^{\dagger}U_1+U_1^{\dagger}U_0\big)\rho\big\},\\
\langle\sigma_y^p\rangle=&\frac{1}{2i}{\rm Tr}\big\{\big(U_0^{\dagger}U_1-U_1^{\dagger}U_0\big)\rho\big\},
\end{aligned}
\end{equation}
and $\langle\sigma_z^p\rangle=0$. As we see, the probe-measurement outcomes $\langle\sigma_x^p\rangle$ and $\langle\sigma_y^p\rangle$ are, respectively, equal to the expectation value of the real and imaginary part of the operator $U_0^{\dagger}U_1$ in the initial dark-system state $\rho$. These expectation values are the information we can extract through Pauli measurements on the probe and by changing the pulse-sequence parameters $\tau$ and $N$ we can control to which dark-system quantity they correspond. The information is extracted by measuring the coherence of the probe and its dephasing thus affects the reconstruction fidelity. The scheme is feasible as long as the extended probe coherence time by the pulsed strategy is longer than the total measurement time. Up to this point we make no assumption on the nature of the dark system.  In the following we give two explicit examples, one discrete and one continuous variable system, and demonstrate in both cases how the unitary $U_0^\dagger U_1$ can be engineered in order to perform a state reconstruction of these dark systems.

\textit{State reconstruction of a spin-1/2 system.---}As a first case we consider a dark spin-1/2 system. Its density operator can be written as $\rho=(\mathbb{1}+\mathbf{r}\cdot\boldsymbol{\sigma})/2$, with the unity operator $\mathbb{1}$, the dark-spin Pauli vector $\boldsymbol{\sigma}$, and the Bloch vector $\mathbf{r}$ fulfilling $\mathbf{r}={\rm Tr}\{\boldsymbol{\sigma}\rho\}$. On the other hand, the unitary $U_0^\dagger U_1$ takes the form $U_0^{\dagger}U_1 =\cos\phi\,\mathbb{1}-i \sin\phi\,\mathbf{n}\cdot\boldsymbol{\sigma}$, with a unit vector $\mathbf{n}$. Comparing this expression with Eq.~\eqref{eq:sigma_p} reveals $\langle\sigma_x^p\rangle=\cos\phi$ and allows us to connect the probe-measurement outcome  $\langle\sigma_y^p\rangle$ with the dark-spin Bloch vector $\mathbf{r}$ according to
\begin{equation}
\label{eq:ex_val}
\langle\sigma_y^p\rangle=-\sin\phi\,\mathbf{n}\cdot\mathbf{r}.
\end{equation}
Three independent measurements for different pulse-sequence parameters $\tau$ and $N$ are thereby  sufficient for a full state reconstruction of the dark spin. It can also be seen that for a faithful measurement of any component of the dark-spin Bloch vector a crucial condition is $\cos\phi=0$. This additionally makes it possible to engineer the pulse sequences such as to obtain the three components $r_\kappa$ separately by ensuring $\sin\phi\,n_\kappa=-1$, for $\kappa=x,y,z$.

\begin{figure}[t]
\includegraphics[width=1\linewidth]{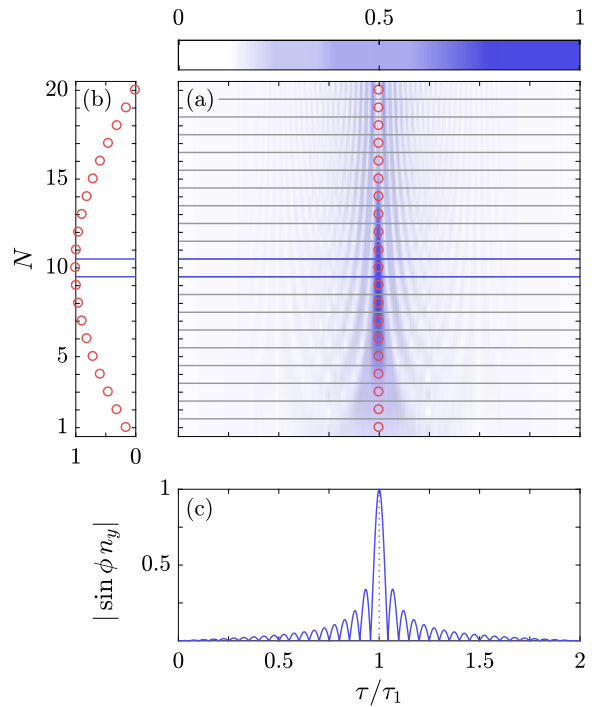}
\caption{\label{fig1} Measurement of $\langle\sigma_y\rangle$ of a dark spin-1/2 system. (a) Dependence of $|\sin\phi\,n_y|$ on the pulse-sequence parameters $\tau$ and $N$ for $a_z=0.015\, \omega_0$ and $a_x=0.08\, \omega_0$. (b) Vertical cut along the free evolution time $\tau_1=2\pi/(\omega_1+\omega_0)$, indicated by red circles in (a), resulting in an optimal pulse-cycle number $N=\pi/4v_x=10$. (c) Horizontal cut along $N=10$, indicated by blue lines in (a).}
\end{figure}

For a general dark-spin Hamiltonian $H_d=(\omega_0/2)\sigma_z$ this reconstruction can be achieved by  the probe-state-dependent potentials $H_0=0$ and $H_1=(a_z/2)\sigma_z+(a_x/2)\sigma_x$, where $a_z$ and $a_x$ arise from the coupling between the probe and the dark spin. This results in the Hamiltonians
\begin{align}
\label{eq:eff_fields_spin}
V_0 =\frac{\omega_0}{2}\sigma_z,\quad V_1 =\frac{\omega_1}{2}\left(v_x\sigma_x+v_z\sigma_z\right),
\end{align}
with $v_x=a_x/\omega_1$, $v_z=(\omega_0+a_z)/\omega_1$, and $\omega_1^2=(\omega_0+a_z)^2+a_x^2$. The above Hamiltonians represent one effective longitudinal field of strength $\omega_0$ associated with the probe ground state and the other one of strength $\omega_1$ associated with the probe excited state, which is tilted from the $z$ direction by the angle $\arctan(v_x/v_z)$. From these state-dependent effective fields and the pulse sequence applied to the probe one can obtain the explicit form of the operator $U_0^\dagger U_1$.

Every pulse-sequence segment of length $2\tau$, i.e., first an evolution under $V_0$ and then under $V_1$, or vice versa, produces a state-dependent rotation given by the unitaries $u_k=\exp(-i\theta \mathbf{n}_k\cdot\boldsymbol{\sigma})$, for $k=0,1$, respectively. Here, the angle $\theta$ satisfies $\cos\theta=\cos\alpha\cos\beta-v_z\sin\alpha\sin\beta$ and the two rotation axes fulfill $\mathbf{n}_0\cdot\mathbf{n}_1=1-2v_x^2\sin^2\alpha\sin^2\beta/\sin^2\theta$, with $\alpha=\omega_0\tau/2$ and $\beta=\omega_1\tau/2$. The operators $U_k=u_k^N$ are then rotations around the same axis, but by the angle $N\theta$ and one obtains the expressions~\cite{Kolkowitz2012,Taminiau2012,Supp}
\begin{gather}
\label{onespin1}
\cos\phi=1-\sin^2(N\theta)(1-\mathbf{n}_0\cdot\mathbf{n}_1),\\
\label{onespin2}
\mathbf{n}=\frac{\sqrt{2(1\!-\!\mathbf{n}_0\!\cdot\!\mathbf{n}_1)}\sin^2(N\theta)}{\sin\phi\sin\theta}
\begin{bmatrix}
\sin\alpha\cos\beta\!+\!v_z\cos\alpha\sin\beta \\
\sin\theta\cot(N\theta) \\
-v_x\cos\alpha\sin\beta
\end{bmatrix}\nonumber
\end{gather}
for the quantities $\phi$ and $\mathbf{n}$. This is their functional dependence on the pulse-sequence parameters $\tau$ and $N$, which can be used to fully determine the Bloch vector $\mathbf{r}$, according to Eq.~\eqref{eq:ex_val}, from three independent probe measurements.

Among the possible choices for the parameters $\tau$ and $N$, which ensure $\cos\phi=0$, we can  choose to measure the observables corresponding to the three components $r_\kappa$ of the Bloch vector, for which the additional condition $\sin\phi\,n_\kappa=-1$ has to be fulfilled, for $\kappa=x,y,z$. As an example, in the $y$ case, these two conditions are $\mathbf{n}_0\cdot\mathbf{n}_1=-1$ and $\sin^2(N\theta)=1/2$. Here, the first one is fulfilled for the evolution time $\tau_1=2\pi/(\omega_1+\omega_0)$ and the second one for $N=\pi/4v_x$, yielding $\sin\phi\,n_y=-1$~\cite{Supp}. The results for the measurement of $r_y$ are illustrated in Fig.~\ref{fig1}, where we show $|\sin\phi\,n_y|$ as a function of $\tau$ and $N$. Our further simulations demonstrate that the measurement protocol is very robust against noise acting on the probe~\cite{Supp}. As a brief note, we mention that without the application of pulses, one would have the unitary $U_0^\dagger U_1=\exp(iV_0\tau) \exp(-iV_1\tau)$ and the reachable points within the Bloch sphere are confined to a cylinder of radius $v_x$ around the $z$ axis, making a measurement of $r_x$ and $r_y$ impossible. The further generalization to multispin dark systems is feasible by employing sufficient controllability conditions and the technique of Cartan decomposition~\cite{Albertini2002}.
We also remark that the measurement of some specific observable may already be of significant interest, e.g., for entanglement and quantum-criticality detection~\cite{Supp,Cai2013,Chen2013,Xu2018}. 

\begin{figure}[t]
\includegraphics[width=\linewidth]{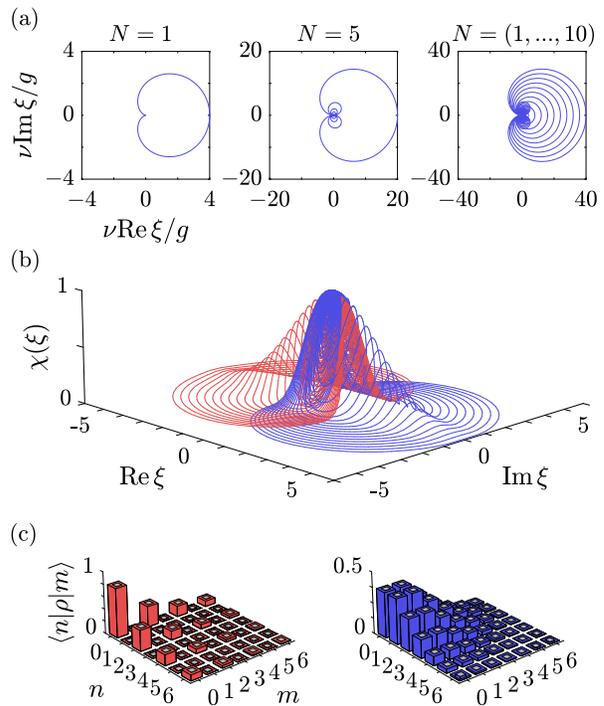}
\caption{\label{fig2} State reconstruction of a dark harmonic oscillator. (a) Reachable points $\xi$ in reciprocal phase space for different numbers of pulse cycles, $N=1$ and $N=5$ in the first two panels and all points for $N=(1,...,10)$ in the right panel. (b) First 20 contour lines (blue) sampled from the characteristic function of a squeezed vacuum state $S(\lambda)|0\rangle$, with $\lambda=\log(1/2)$ and $g/\nu=3/40$. Red curves are obtained using $\chi(-\xi)=\chi(\xi)^\ast$. (c) Density matrices reconstructed from an interpolated characteristic function obtained from $N=(1,...,20)$ for $g/\nu=3/40$. Left panel: Squeezed vacuum from (b). Right panel: Coherent state $|\eta\rangle=D(\eta)|0\rangle$ with $\eta=1$. Gray inner bars represent the exact values for comparison.}
\end{figure}

\textit{State reconstruction of a harmonic oscillator.---}As a second case we consider a continuous-variable dark system which is formed by a harmonic oscillator of frequency $\nu$ with the annihilation operator $a$. The interaction between the probe and the harmonic oscillator is assumed to be of the form $H_{\rm int}=(g/2)\sigma_z^p(a+a^\dagger)$, leading to the state-dependent Hamiltonians
\begin{align}
\label{eq:eff_fields_osc}
V_0=\nu a^\dagger a-\frac{g}{2}(a+a^\dagger),\quad V_1=\nu a^\dagger a+\frac{g}{2}(a+a^\dagger).
\end{align}
Using the multiplication properties of the displacement operator $D(\eta)=\exp(\eta a^\dagger-\eta^\ast a)$, in a picture displaced by $g/2\nu$, the operator $U_0^\dagger$ can be brought into the form $D(\zeta)\exp(2iN\nu\tau a^\dagger a)$, while $U_1$ similarly is of the form $D(\zeta^\ast)\exp(-2iN\nu\tau a^\dagger a)$, where $\zeta$ is a function of $\tau$ and $N$~\cite{Supp}.  In this way, we can obtain the unitary $U_0^\dagger U_1=D(\xi)$ in the simple form of a single displacement operator with the quantity
\begin{gather}
\xi(\tau,N)=-2\frac{g}{\nu}\sin(N\nu\tau)\tan\left(\frac{\nu\tau}{2}\right)e^{iN\nu\tau},
\end{gather}
which depends on the pulse-sequence parameters $\tau$ and $N$~\cite{Supp}. Equation~\eqref{eq:sigma_p} then yields
\begin{align}
\langle\sigma_x^p\rangle+i\langle\sigma_y^p\rangle=\chi\left(\xi\right),
\end{align}
with the Wigner characteristic function $\chi$, which is defined as $\chi(\xi)={\rm Tr}\{D(\xi)\rho\}$~\cite{Cahill1969a}. This function over reciprocal phase space is the complex Fourier transform of the Wigner function~\cite{Wigner1932} and contains all information on the initial density operator $\rho$ of the harmonic oscillator. For a full knowledge of the characteristic function, the completeness of the displacement operators~\cite{Cahill1969a} allows an exact reconstruction of the density operator itself according to $\rho=\int{\rm d}^2\xi\,\chi(\xi)D^\dagger(\xi)/\pi$. For example, in the Fock basis, the matrix elements $\langle n|\rho|m\rangle$ can easily be obtained using this expression and the matrix elements of the displacement operator $\langle n|D^\dagger(\xi)|m\rangle$~\cite{Cahill1969a}.

In our scheme, every fixed pulse-sequence parameter $N$ corresponds to a closed curve $\xi(\tau,N)$ in reciprocal phase space, shown in Fig.~\ref{fig2}(a) for several pulse numbers. Their periodicity in $\tau$ is determined by the harmonic oscillator frequency $\nu$, requiring a maximal necessary evolution time of $\tau=2\pi/\nu$. The maximal distance from the origin is reached for $\tau=\pi/\nu$ and has the value  $4Ng/\nu$; i.e., it scales linearly with the pulse number. By varying $N$ we can sample the characteristic function along this manifold of curves, as shown by blue lines in Fig.~\ref{fig2}(b) for the example of a squeezed vacuum state $S(\lambda)|0\rangle$, with $S(\lambda)=\exp[(\lambda^\ast a^2-\lambda a^{\dagger 2})/2]$. For a real squeezing parameter $\lambda=\log(1/2)$, as chosen here, and $\xi=\xi_{\rm r}+i\xi_{\rm i}$ the corresponding characteristic function is $\chi(\xi)=\exp(-\xi_{\rm r}^2/8-2\xi_{\rm i}^2)$. The property $\chi(-\xi)=\chi(\xi)^\ast$ allows us to obtain the values of $\chi$ along these curves mirrored around the origin by complex conjugation, as represented by red curves in Fig.~\ref{fig2}(b).

Figure~\ref{fig2}(c) shows the density matrices reconstructed from the characteristic function for the squeezed vacuum state from Fig.~\ref{fig2}(b) and a coherent state by an interpolation of $\chi$ using $N=(1,...,20)$. The results are in good agreement with the exact density matrices,  showing trace distances of the order $10^{-3}$~\cite{Supp}. As a comparison, for the case of no pulses applied to the probe only points on the circle $\xi=g[\exp(i\nu\tau)-1]/\nu$ in reciprocal phase space can be sampled, which would be insufficient for a satisfactory state reconstruction.

The fact that the characteristic function has its maximum $\chi(0)=1$ at the origin, and is mostly centered in this region, is favorable in experiments since the density of reachable points $\xi$ is high close to the origin. In contrast, the Wigner function, which contains the same information, can have its maximum at any point in phase space, making it necessary to scan over larger areas with schemes for its measurement. We also stress that contrary to other methods for the measurement of phase-space distributions, this scheme does not require any manipulation of the harmonic oscillator, such as a displacement operation prior to the measurement procedure, or a control of the coupling strength. This advantage would become particularly important for systems in which direct manipulation on the harmonic oscillator is hard to achieve.	

\textit{Potential experimental implementations.---}As an example for the state tomography of a spin-1/2 system, we use a nitrogen-vacancy (NV) center as a probe and a dark spin of a weakly coupled $^{13}$C nucleus in diamond~\cite{Doherty2013}. Under an external magnetic field of strength $B$ along the NV axis, i.e., the $z$ direction, the Hamiltonians then read $H_{\rm NV}=D S_z^2+\gamma_eBS_z$ and $H_{\rm C}=\gamma_{\rm C}BI_z$, where $\gamma_e$ and $\gamma_{\rm C}$ are the gyromagnetic ratios of the NV-center spin and the $^{13}$C nuclear spin, respectively, and $D/2\pi=2.87$~GHz is the electron-spin zero-field splitting. Furthermore, the components of $S_\kappa$ and $I_\kappa$ denote their respective spin-$1$ and spins-$1/2$ operators, for $\kappa=x,y,z$. One can choose the $x$ direction such that the hyperfine interaction between the NV center and the nuclear spin is of the form $H_{\rm hf}=A_{\parallel}S_z I_z+A_{\perp}S_z I_x$~\cite{Taminiau2012}. The external magnetic field allows us to address specific transitions of the NV-center electronic states and thereby, for example, to use the two states $|0\rangle_p=|m_s\!=\!0\rangle$ and $|1\rangle_p=|m_s\!=\!1\rangle$ as our probe. This results in the state-dependent effective fields acting on the $^{13}$C nuclear spin given by Eq.~\eqref{eq:eff_fields_spin} with $\omega_0=\gamma_{\rm C} B$, $a_z=A_\parallel$, $a_x=A_\perp$, and $\sigma_\kappa=2I_\kappa$, for $\kappa=x,y,z$. As an example, we consider a weakly coupled $^{13}$C with $A_\parallel/2\pi=2.54$ and $A_\perp/2\pi=13.22$~kHz under a magnetic field $B=15.4$~mT, which are the parameters used in Fig.~\ref{fig1}. The assumption of instantaneous $\pi$ pulses is well justified, since pulse durations $t_\pi$ of tens of nanoseconds have been realized, and the free-evolution time for the measurement of Bloch vector components thereby fulfill $t_\pi\ll\tau$~\cite{de-Lange2010,Robledo2010,de-Lange2011}. The condition $2N\tau<T_{2p}$ for the total evolution time can also be satisfied for achievable long probe coherence times $T_{2p}$~\cite{Ryan2010,Naydenov2011,Bar-Gill2013}.

To show the feasibility of an experimental demonstration for a continuous-variable dark system, we consider the motional-state reconstruction of a single trapped ion in a magnetic field gradient~\cite{Johanning2009,Khromova2012}. We orient ourselves at parameters from Ref.~\cite{Sriarunothai2017} with single $^{171}$Yb$^+$ ions trapped in a linear Paul trap with an axial frequency $\nu/2\pi=117$~kHz. In this setup, a magnetic field of the form $B(z)=B_0+B_1 z$ is applied along the trap axis $z$. As a probe two-level system one can choose the two sublevels $|0\rangle_p=|m_s\!=\!-1/2\rangle$ and $|1\rangle_p=|m_s\!=\!1/2\rangle$ with $m_F=0$ of the $^2S_{1/2}$ state, whose coherence time can be longer than 1000 s~\cite{Fisk1997}. The linear magnetic field gradient $B_1$ induces a coupling between the ion motion and the probe of the form $H_{\rm int}=(g/2)\sigma_z^p(a+a^\dagger)$, yielding the Hamiltonians $V_0$ and $V_1$ given in Eq.~\eqref{eq:eff_fields_osc}. The coupling strength is determined by $g=2\mu_{\rm B} B_1 /\sqrt{2M\hbar\nu}$, with the Bohr magneton $\mu_{\rm B}$ and the ion mass $M$. For the reported magnetic field gradient $B_1=19$~T/m, this results in a coupling ratio of $g/\nu=0.072$, which is roughly the one we used above in Fig.~\ref{fig2}. In this system, high-fidelity $\pi$ pulses with durations on the order of tens of picoseconds have also been demonstrated for these $^2S_{1/2}$ states~\cite{Campbell2010}.

\textit{Conclusions and outlook.---}We present a general scheme for the quantum-state reconstruction of a dark system, which is inaccessible for direct control and measurements. The scheme only relies on the pulsed control and readout of a probe-two-level system, while requiring no manipulation of the target system. We illustrate our idea at the state tomography of a spin and a harmonic oscillator. For both examples, we show the feasibility to implement proof-of-principle demonstrations in currently available experimental setups. Moreover, the measurement scheme is intrinsically robust against slow noise acting on the probe due to the incorporated dynamic decoupling. The present idea provides a versatile tool for quantum-state measurement and can be extended to more general scenarios, such as dark systems formed by higher spins, many-body systems, and novel mechanical systems. A further generalization to a sequence of nonequidistant pulses and continuous processes is possible and may increase the information that can be obtained of the dark system.

The authors thank C. Arenz, Y. Chu, S. B. J\"ager, G. Morigi, M. B. Plenio, and R. Said for helpful discussions and comments. This work was supported by the National Natural Science Foundation of China (Grants No. 11574103, No. 11874024, No. 11690030, and No. 11690032), the National Key R\&D Program of China (Grant No. 2018YFA0306600), the China Postdoctoral Science Foundation (Grant No. 2017M622398), and the National Young 1000 Talents Plan.

\clearpage

\setcounter{equation}{0}
\setcounter{figure}{0}
\setcounter{table}{0}
\setcounter{page}{1}
\makeatletter
\renewcommand{\thesection}{S.\arabic{section}}
\renewcommand{\thesubsection}{\thesection.\arabic{subsection}}
\renewcommand{\theequation}{S\arabic{equation}}
\renewcommand{\thefigure}{S\arabic{figure}}
\renewcommand{\bibnumfmt}[1]{[S#1]}
\renewcommand{\citenumfont}[1]{S#1}

\onecolumngrid
\begin{center}
\textbf{\large Supplemental Material: \\
Pulsed Quantum-State Reconstruction of Dark Systems}
\end{center}
\twocolumngrid

\section{S.1. Spin-$\frac{1}{2}$ system}	
\subsection{S.1.1. Basic principle}
For the pulse sequence applied to the probe in the presented state-reconstruction scheme a single pulse-sequence segment, which is applied $N$ times, has the form $[\pi\!-\!\tau\!-\!\pi\!-\!\tau]$, i.e., an evolution time $\tau$, a $\pi$ pulse, a second evolution time $\tau$ followed by a final $\pi$ pulse. For such a segment the probe-state-dependent evolution operators for a dark spin 1/2 read
\begin{eqnarray}
	\begin{aligned}
		u_0=&e^{-i\beta(v_z\sigma_z+ v_x\sigma_x)} e^{-i\alpha\sigma_z},\\
		u_1=&e^{-i\alpha\sigma_z} e^{-i\beta(v_z\sigma_z+v_x\sigma_x)},
	\end{aligned}
\end{eqnarray}
where $v_x=a_x/\omega_1$, $v_z=(\omega_0+a_z)/\omega_1$, $\omega_1^2=(\omega_0+a_z)^2+a_x^2$, $\alpha=\omega_0 \tau/2$ and $\beta=\omega_1 \tau/2$. Since these operators are composed of two rotations, they can both be combined into a single rotation $u_k=\exp(-i\theta\textbf{n}_k\cdot\boldsymbol{\sigma})$, with $k=0,1$, where the angle fulfills $\cos\theta=\cos\alpha\cos\beta-v_z\sin\alpha\sin\beta$, and the two rotation axes are
\begin{align}
	\label{supp:u01}
	\textbf{n}_0=\frac{1}{\sin\theta}\begin{bmatrix}
		v_x\cos\alpha\sin\beta \\
		v_x\sin\alpha\sin\beta \\
		\sin\alpha\cos\beta+v_z\cos\alpha\sin\beta
	\end{bmatrix}
\end{align}
and $\mathbf{n}_1=(n_0^x,-n_0^y,n_0^z)$. For the full pulse sequence we can thereby write $U_k=\exp(-iN\theta\,\textbf{n}_k\cdot\boldsymbol{\sigma})$~\cite{sKolkowitz2012,sTaminiau2012}, for $k=0,1$, and the unitary $U_0^\dagger U_1$ can likewise be written as a single rotation $U_0^{\dagger}U_1=\exp(-i\phi\,\textbf{n}\cdot\boldsymbol{\sigma})$, with
\begin{gather}
	\label{supp:phi-n}
	\cos\phi=\cos^2(N\theta)+\sin^2(N\theta) (\textbf{n}_0\cdot\textbf{n}_1),\\
	\textbf{n}=\frac{\sin(N\theta)}{\sin\phi}\left[\cos(N\theta)(\textbf{n}_1-\textbf{n}_0)
	-\sin(N\theta)(\textbf{n}_0\times\textbf{n}_1)\right].\nonumber
\end{gather}
Using $\textbf{n}_0\cdot\textbf{n}_1=1-2v_x^2\sin^2\alpha\sin^2\beta/\sin^2\theta$ this corresponds to the expressions given in the main text. On the other hand, for a time evolution of time $\tau$ without the application of pulses the unitary corresponding to $U_0^\dagger U_1$ merely has
the form of $u_1$ with $\alpha$ replaced by $-\alpha$.

\begin{figure}[t]
	\centering
	\includegraphics[width=\linewidth]{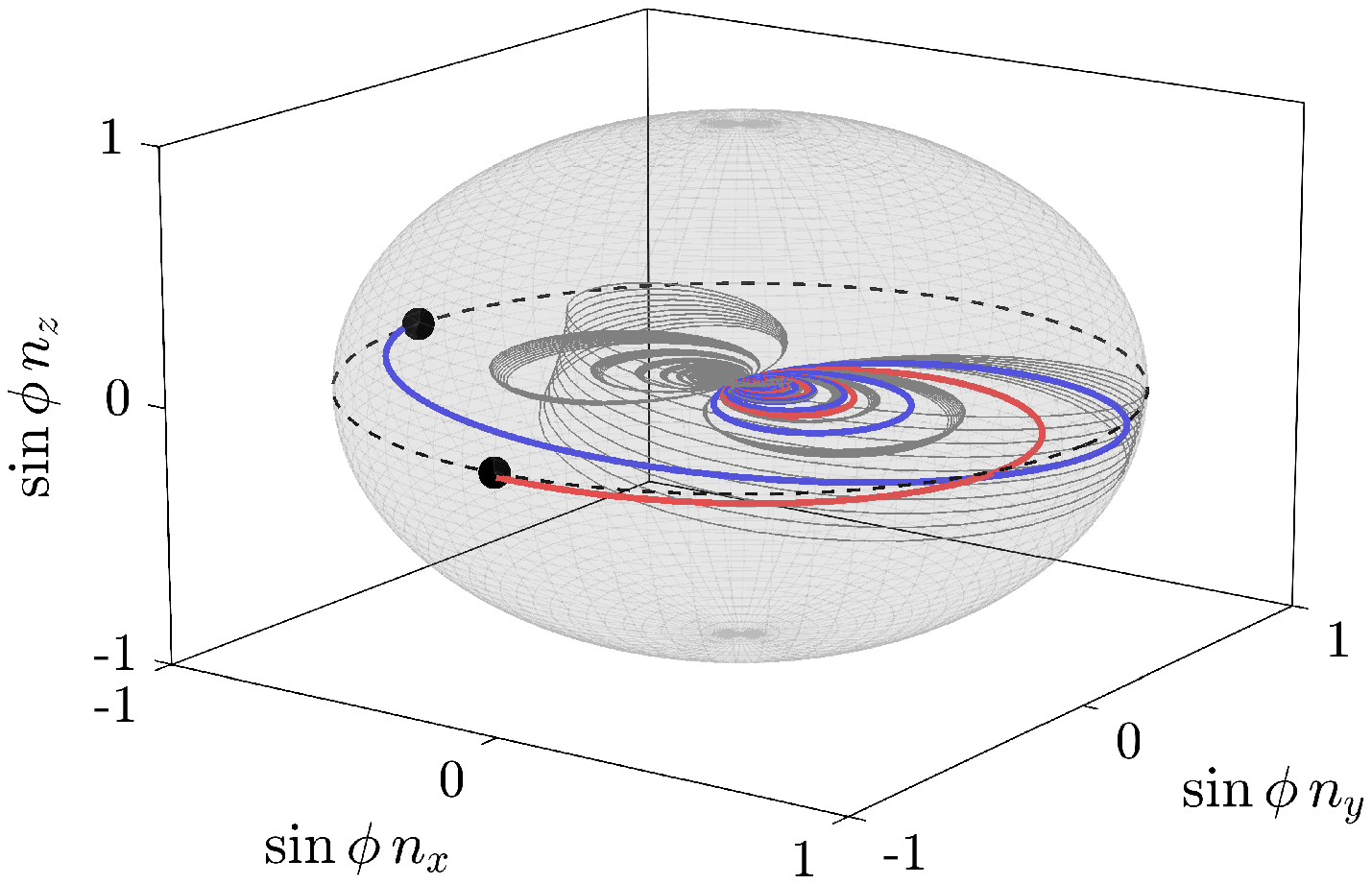}
	\caption{\label{figs1}Trajectories of $\sin\phi\,\mathbf{n}$ when sweeping over $\tau$ for fixed $N$. Red: $N=10$, $\tau\in[0,\tau_1]$. Blue: $N=14$, $\tau\in[0,1.02\,\tau_1]$. Gray: $N=20$, $\tau\in[0,20\,\tau_1]$. Black dots represent measurement conditions for the Bloch vector components of $r_x$ and $r_y$, the dashed line indicates the equator.}
\end{figure}

As an example to illustrate our idea, we now consider measurements of a single components of the Bloch vector $\mathbf{r}$, under the typical situation of weak coupling, viz. $a_x,a_z\ll\omega_0$. In this case we find $v_z=1-v_x^2/2$. And in first order of $v_x$ the $z$ component $\sin\phi\,n_z$ can be neglected, which is valid for relatively short evolution times $\tau$. This leads to the measurement condition for $x$ and $y$, namely the proper parameters $\tau$ and $N$, given in the main text: (i) $n_{\kappa}=-1$ with $\kappa=x,y$ and (ii) $\cos\phi=0$.
First we focus on the $y$ component, where condition (i) implies $n_x=0$, which means $\sin\alpha\cos\beta+v_z\cos\alpha\sin\beta=0$. In first order of $v_x$ this leads to the condition $\tau_k=2k\pi/(\omega_1+\omega_0)$, with odd integers $k$. One also has $\theta=v_x$ and $\sin\alpha\sin\beta=1$, resulting in two anti-parallel rotation axes, viz. $\textbf{n}_0\cdot\textbf{n}_1=-1$. For condition (ii) this yields $\cos\phi=\cos(2N\theta)\approx\cos(2N v_x)=0$ and thereby $N=(l+\pi/2)/2v_x$. Here we choose $k=1$ and $l=0$ to minimize the total measurement time. In summary, the $r_y$-measurement conditions are
\begin{equation}
	\label{ycondition}
	\tau_1=\frac{2\pi}{\omega_1+\omega_0},\quad N=\frac{\pi}{4v_x}.
\end{equation}
Similarly, in the $x$ case, conditions (i) and (ii) read $n_y=0$ and $\cos(N\theta)=0$, which is fulfilled for perpendicular rotation axes, $\textbf{n}_1\cdot\textbf{n}_0=0$. Figure~\ref{figs1} shows three examples of the points $\sin\phi\,\mathbf{n}$ for different pulse-segment numbers (red, blue, and gray).

\subsection{S.1.2. Measurement of unknown coupling constants}
A measurement of the probe-expectation value $\langle\sigma_x^p\rangle$, which is equal to $\cos\phi$, includes the possibility to extract the coupling constants $a_z$ and $a_x$ in case they are unknown~\cite{sGrinolds2014,sZopes2018}. For every value of $N$ the function $\cos\phi$ shows a minimum at the free-evolution time $\tau_1=2\pi/(\omega_1+\omega_0)$, which means in principle a single swipe over $\tau$, for a fixed $N$, is sufficient to determine $\tau_1$. In a second step the pulse number can be varied for the fixed evolution time $\tau_1$ in order to find the $N$ that minimizes $|\cos\phi|$. As in the case for a measurement of the $y$ component this is fulfilled for the integer closest to $\pi/4|v_x|$.
\begin{figure}[t]
	\includegraphics[width=\linewidth]{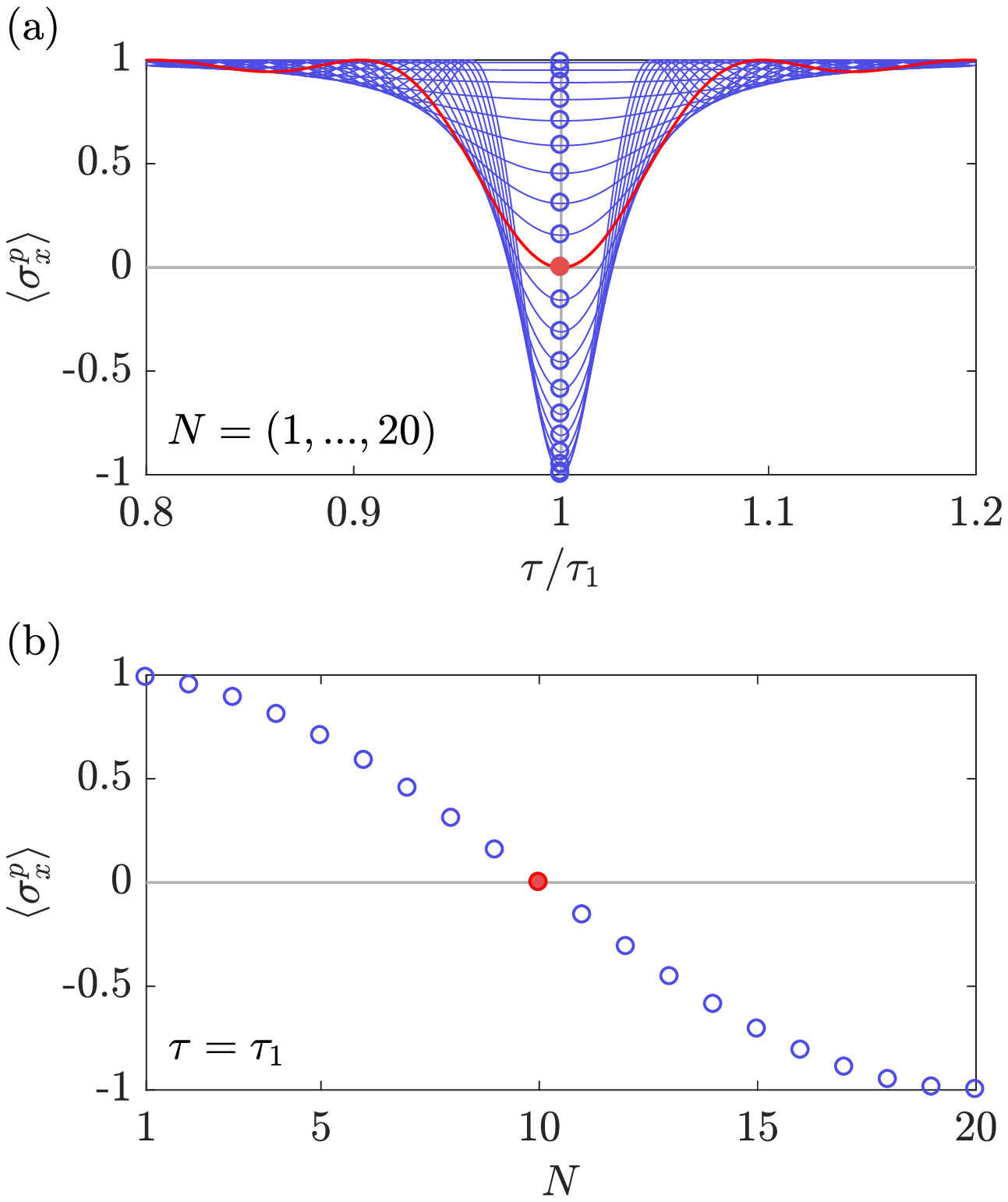}
	\caption{\label{figs2}Procedure for the measurement of the coupling constants $a_z$ and $a_x$. (a) Measurement of $\langle\sigma_x^p\rangle=\cos\phi$, from which $\tau_1=2\pi/(\omega_1+\omega_0)$ can be extracted as the time where the curves take their minima (circles). (b) Value of $\langle\sigma_x^p\rangle$ at $\tau_1$, determined by the measurements shown in (a) used to obtain the optimal pulse number $N$. The ideal value $N=\pi/4v_x$  is indicated by the red line in (a) and the red circle in (b).}
\end{figure}
This procedure is illustrated in Fig~\ref{figs2} for the values from the main text, viz. $a_z=0.015\,\omega_0$ and $a_x=0.08\,\omega_0$. An experimental determination of $\tau_1$ and the optimal $N$ allows to infer the coupling constants $a_z$ and $a_x$ in dependence of the known $\omega_0$. These may then be used to locate the dark spin if, for example, the coupling is given by a magnetic dipole-dipole interaction.

\subsection{S.1.3. Robustness against noise}
\begin{figure}[t]
	\centering
	\includegraphics[width=\linewidth]{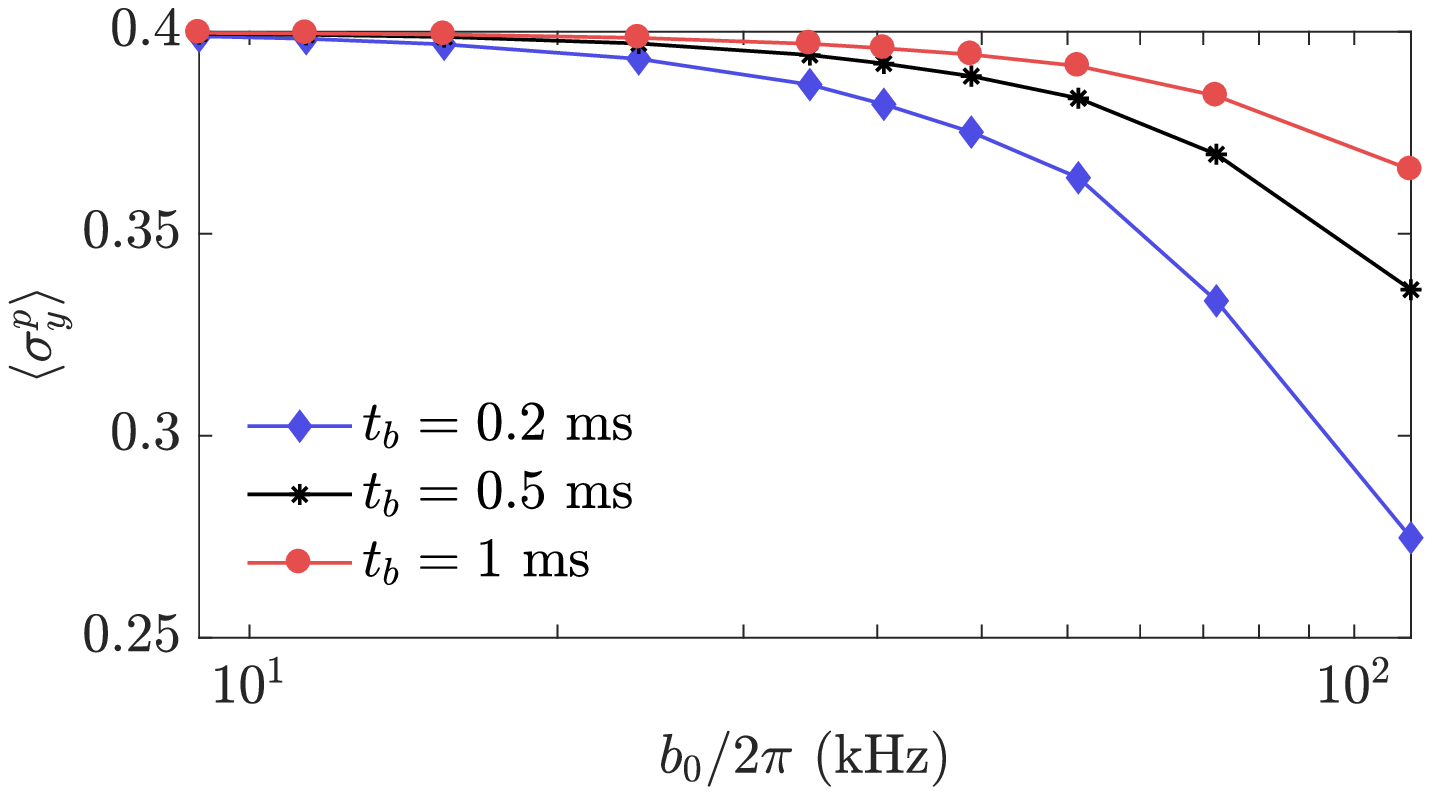}
	\caption{\label{figs3}Simulation of the $\langle\sigma_y\rangle$ measurement of the nuclear spin under the influence of noise on the probe in dependence on the noise variance $b_0$ for three different noise-correlation times $t_b$. The parameters are from the main text with $N=10$ and $\tau=3\ \mu$s.}
\end{figure}

In this section, we demonstrate the robustness of the proposed scheme against noise acting on the probe. We exemplify its robustness at the experimental implementation for the state tomography of a $^{13}$C nuclear spin weakly coupled to an NV-center probe. The main source of noise in this solid-state system comes from the spin bath surrounding the NV center. An established description of the collective effect of this complex environment on the probe is that of a fluctuating magnetic field~\cite{sDobrovitski2008,sDobrovitski2009,sde-Lange2010,sMaze2012,sWang2013}, resulting in the noise Hamiltonian $H_{\rm noise}=(b(t)/2)\sigma_z^p$. Here, the noise $b(t)$ is a random variable
obeying a zero-mean Gaussian distribution with the autocorrelation $\langle b(t)b(0)\rangle=b_0^2\exp(-t/t_b)$, where $b_0^2$ denotes the variance and $t_b$ the correlation time of the noise. The time evolution of this random variable can be modelled as an Ornstein-Uhlenbeck process~\cite{sVanKampen2007} with the update formula as follows~\cite{sGillespie1991,sGillespie1995,sGillespie1996}
\begin{equation}
	b(t+\Delta t)=b(t)e^{-\Delta t/t_b}+b_0\sqrt{1-e^{-2\Delta t/t_b}}\mathcal{N},
\end{equation}
with a normally distributed random variable $\mathcal{N}$ and the time discretization $\Delta t$.
Duo to the symmetrical structure of the pulse sequence, our scheme will be robust against slow noise, i.e., noise with a correlation time $t_b$ which is long compared to the free evolution time $\tau$ between pulses. This is evident in Fig.~\ref{figs3}, where we show simulations for a $\langle\sigma_y\rangle$ measurement of the dark spin under the influence of the noise on the probe. Here, we use the parameters given in the main text and perform simulations for the three typical noise correlation times $t_b=(0.2,0.5,1)$~ms~\cite{sDobrovitski2008,sDobrovitski2009,sde-Lange2010,sMaze2012,sWang2013} and different noise variances $b_0/2\pi\in[9,112]$ kHz, which are related to the probe coherence time $T_{2p}^\ast$, as measured in free-induction decay experiments, via $T_{2p}^\ast=\sqrt{2}/b_0$~\cite{sDobrovitski2008,sDobrovitski2009,sde-Lange2010,sMaze2012,sWang2013}. In our simulations, we assume the $^{13}$C spin to be initially in the state $\rho=(\mathbb{1}+\mathbf{r}\cdot\boldsymbol{\sigma})/2$ with $\mathbf{r}=(0,0.4,0)$ and average the random process over 1000 realizations. The parameters are the ones from the main text, $N=10$ and $\tau=3\ \mu$s, leading to a total measurement time of $60\ \mu$s, which lies well below probe coherence times $T_{2p}$ achieved with pulsed dynamical decoupling~\cite{sde-Lange2010,sNaydenov2011}.

\subsection{S.1.4. Two interacting spin-$\frac{1}{2}$}
As a further example for the applicability of the proposed scheme, we consider a dark system formed by two  spin-1/2, whose Hamiltonian is given by
\begin{equation}
	H_d=\sum_{k=1,2}\frac{\omega_0}{2} \sigma_z^{(k)}+\frac{A_x}{2}\sigma_x^{(1)}\sigma_x^{(2)},
\end{equation}
with the Pauli operators $\sigma_\lambda^{(k)}$, for $k=1,2$ and $\lambda=x,y,z$, and the inter-dark-spin coupling strength $A_x$. Analogous to the single-spin case in the main text, the probe-state-dependent potentials are assumed to be $H_0=0$ and 
\begin{equation}
	H_1=\sum_{k=1,2}\bigg[\frac{a_z^{(k)}}{2}\sigma_z^{(k)}+\frac{a_x^{(k)}}{2}\sigma_x^{(k)}\bigg].
\end{equation}
If we now consider the weak-coupling regime, where $a_z^{(k)},a_x^{(k)},A_x\ll\omega_0$ is fulfilled, the transitions in the manifold $\{|0\rangle_1|0\rangle_2,|1\rangle_1|1\rangle_2\}$ of the bipartite dark system can be neglected due to their large energy separation. We therefore focus on the subspace spanned by the two states $|\tilde 0\rangle=|0\rangle_1|1\rangle_2$ and $|\tilde 1\rangle=|1\rangle_1|0\rangle_2$, which can be considered as a pseudo spin with the Pauli vector $\tilde{\boldsymbol{\sigma}}$. In this subspace, the resulting state-dependent potentials, in the interaction picture with respect to the probe Hamiltonian, are then simply given by
\begin{equation}
	V_0=\frac{A_x}{2}\tilde\sigma_x,\quad V_1=\frac{A_x}{2}\tilde\sigma_x+\frac{A_z}{2}\tilde\sigma_z,
\end{equation}
with $A_z=a_z^{(1)}-a_z^{(2)}$. For the probe-measurement outcome we find  $\langle\sigma_y^p\rangle=-\sin\phi\,\tilde{\mathbf{n}}\cdot\langle\boldsymbol{\tilde{\sigma}}\rangle$, where we can apply Eq.~(5) of the main text, with $\tilde{\mathbf{n}}=(n_z,n_y,n_x)$ and the substitutions $\alpha=A_x\tau/2$, $\beta=A\tau/2$, $A^2=A_x^2+A_z^2$, $v_x=A_z/A$, and $v_z=A_x/A$. The Bloch vector $\tilde{\mathbf{r}}=\langle\boldsymbol{\tilde{\sigma}}\rangle$ of the pseudo spin in this subspace can be determined by varying the pulse-sequence parameters $\tau$ and $N$, as described in the main text. Our strategy thereby yields a feasible method to measure correlations between the two dark spins, since the Bloch vector components $\tilde{r}_x$ and $\tilde{r}_y$ correspond to
\begin{eqnarray}
	\begin{aligned}
	\tilde{r}_x=\frac{1}{2}\langle\sigma_x^{(1)}\sigma_x^{(2)}+\sigma_y^{(1)}\sigma_y^{(2)}\rangle,\\
	\tilde{r}_y=\frac{1}{2}\langle\sigma_y^{(1)}\sigma_x^{(2)}-\sigma_x^{(1)}\sigma_y^{(2)}\rangle,
	\end{aligned}
\end{eqnarray}
which may be used as entanglement witnesses. 

\section{S.2. Harmonic oscillator}

\subsection{S.2.1. Basic principle}

For a dark harmonic oscillator we show the derivation of the explicit form of the the operator $U_0^\dagger$, the operator $U_1$ can be derived along the same lines. Defining $\epsilon=g/2\nu$ the probe-state-dependent potentials $V_0$ and $V_1$ can be displaced according to
\begin{equation}
	D^\dagger(\epsilon)V_0D(\epsilon)=D(\epsilon)V_1D^\dagger(\epsilon)=\nu a^\dagger a-\epsilon^2\nu.
\end{equation}
The multiplication property of displacement operators, $D(x)D(y)=\exp(i\,{\rm Im}\,xy^\ast)D(x+y)$, and the identity $\exp(ix a^\dagger a)D(y)\exp(-ix a^\dagger a)=D(y\exp(ix))$ allows to write the single pulse-segment evolution operator $u_0$ in the form
\begin{equation}
	u_0=e^{i\varphi_1}D^\dagger(\epsilon)e^{-2i\nu\tau a^\dagger a}D\left(2\epsilon e^{i\nu\tau}\right)D^\dagger(\epsilon),
\end{equation}
with $\varphi_1=2\epsilon^2\nu\tau$. By Hermitian conjugating and taking the operator to the $N$th power we can write $U_0^\dagger=(u_0^\dagger)^N$ as
\begin{equation}
	U_0^\dagger=e^{i\varphi_2}D^\dagger(\epsilon)\left[D\left(2\epsilon \left(1-e^{i\nu\tau}\right)\right)e^{2i\nu\tau a^\dagger a}\right]^ND(\epsilon)
\end{equation}
where $\varphi_2$ incorporates $-N\varphi_1$ and an additional phase that we abstain from writing explicitly, since the operator $U_1$ carries exactly the opposite one. In a next step we rewrite the powers using the identities
\begin{eqnarray}
	\label{supp:displ_identity}
	\begin{aligned}
		\left[D(x)e^{iy a^\dagger a}\right]^N&=\left[\prod_{n=0}^{N-1}D\left(x e^{iny}\right)\right]e^{iNya^\dagger a}\\
		&=e^{i\vartheta}D\left(x\frac{1-e^{iNy}}{1-e^{iy}}\right)e^{iNya^\dagger a},
	\end{aligned}
\end{eqnarray}
where we also do not need to determine the phase $\vartheta$. In these identities, the first equality can be easily proven by induction and the second one follows from the multiplication property of the displacement operator and the geometric sum. This leaves us with
\begin{eqnarray}
	\begin{aligned}
		U_0^\dagger=e^{i\varphi_3}D^\dagger(\epsilon)D(&\zeta)e^{2iN\nu\tau a^\dagger a}D(\epsilon),\\
		U_1=e^{-i\varphi_3}D^\dagger(\epsilon)D(&\zeta^\ast)e^{-2iN\nu\tau a^\dagger a}D(\epsilon),
	\end{aligned}
\end{eqnarray}
where $\varphi_3$ includes $\varphi_2$ and the phase corresponding to $\vartheta$ from identity~\ref{supp:displ_identity}. The quantity $\zeta$ mentioned in the main text thereby has the form
\begin{equation}
	\zeta=\frac{g}{\nu}\left(1-e^{i\nu\tau}\right)\frac{1-e^{2iN\nu\tau}}{1-e^{2i\nu\tau}}
\end{equation}
and the relevant unitary $U_0^\dagger U_1$ finally reads
\begin{equation}
	U_0^\dagger U_1=D\left(-2\frac{g}{\nu} \sin(N\nu\tau)\tan\left(\frac{\nu\tau}{2}\right)e^{iN\nu\tau}\right),
\end{equation}
where the arguments of the displacement operator are the curves $\xi(\tau,N)$.

\subsection{S.2.2. Reconstruction example}
In the Fock basis, the matrix elements $\langle n|\rho|m\rangle$ of a density operator $\rho$, with the associated characteristic function $\chi(\xi)={\rm Tr}\{D(\xi)\rho\}$, can be written as
\begin{equation}
	\label{matrix_elements}
	\langle n|\rho|m\rangle=\frac{1}{\pi}\int d^2\xi\,\langle n|D(-\xi)|m\rangle\chi(\xi).
\end{equation}
In order to calculate their explicit value, one can then use the form of the displacement-operator matrix elements~\cite{sCahill1969a}, which is given by
\begin{equation}
	\label{matrix_elements_D}
	\langle n|D(\eta)|m\rangle=\sqrt{\frac{m!}{n!}}\eta^{n-m}L_m^{(n-m)}\left(|\eta|^2\right)e^{-\frac{|\eta|^2}{2}},
\end{equation}
with the generalized Laguerre polynomials $L$. This expression is valid for $n\geq m$, while for $n<m$ one has to replace $\eta$ by $-\eta^\ast$ as well as exchange $n$ and $m$ on the right-hand side of Eq.~\eqref{matrix_elements_D}.

\begin{figure}[t]
	\centering
	\includegraphics[width=\linewidth]{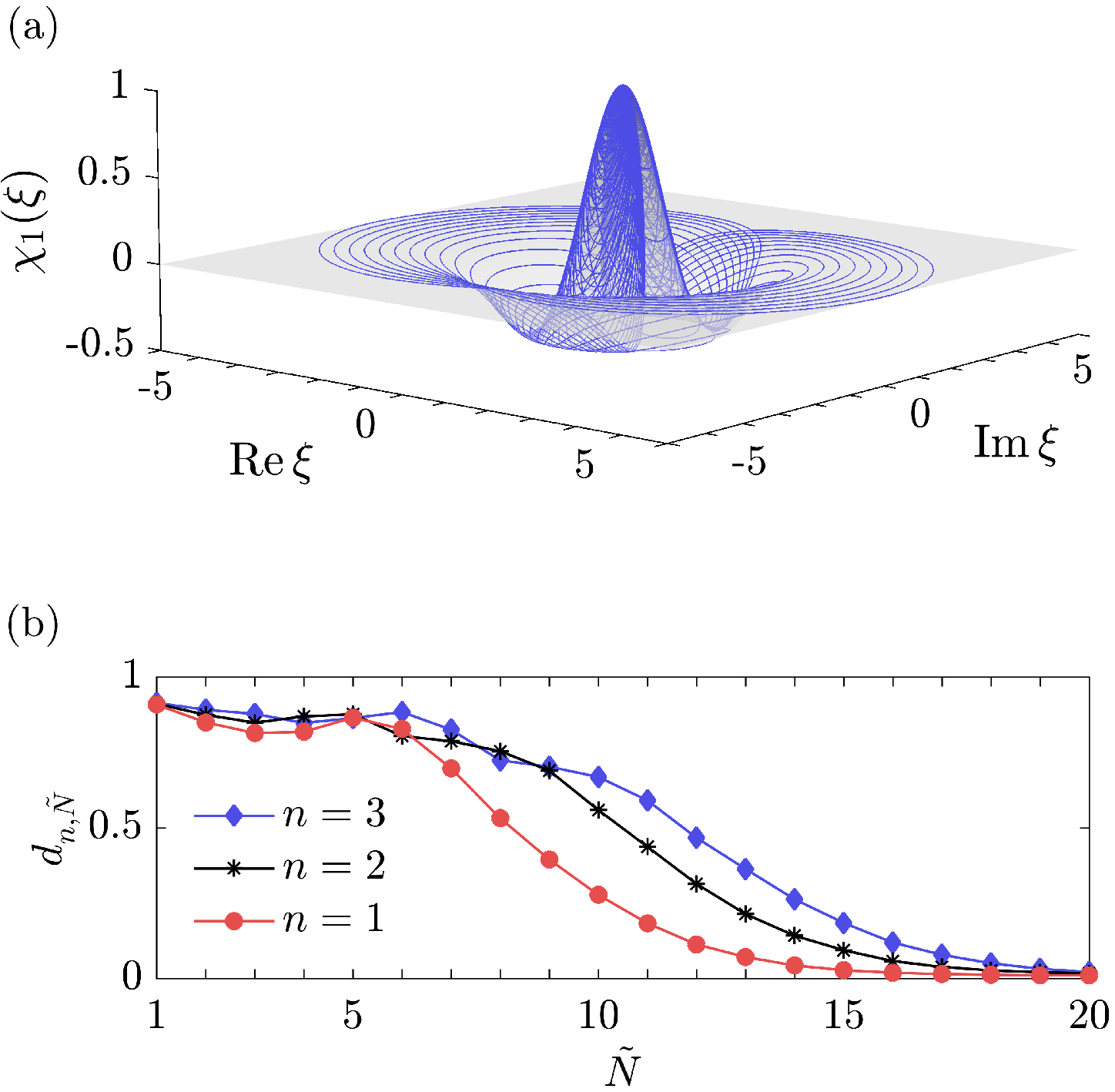}
	\caption{\label{figs4}Reconstruction of a Fock state $|n\rangle$. (a) Exact characteristic function $\chi_1(\xi)$ [gray] and the contour lines $\xi(\tau,N)$ [blue] for $N=(1,...,20)$, $\tau\in[0,2\pi/\nu]$, and $g/\nu=3/40$. (b) Trace distance $d_{n,\tilde{N}}$ between the exact density operator $|n\rangle\langle n|$ and the state $\rho_{n,\tilde{N}}$ reconstructed using an interpolation over the contour lines with $N\leq\tilde{N}$.}
\end{figure}

Here, we demonstrate the state reconstruction using the matrix elements from Eq.~\eqref{matrix_elements} obtained from an interpolated characteristic function. As an example, we assume a dark harmonic oscillator initially prepared in the Fock state $|n\rangle$. The characteristic function of this state has the form $\chi_n(\xi)=L_n^{(0)}\left(|\xi|^2\right)\exp(-|\xi|^2/2)$, as can be see from Eq.~\eqref{matrix_elements_D} for $n=m$. In Fig.~\ref{figs4}(a) we show $\chi_1(\xi)$ [gray] with the contour lines $\xi(\tau,N)$ [blue] for $N=(1,...,20)$ and $g/\nu=3/40$. In our case $\chi(\xi)$ can be interpolated from samples of the exact characteristic functions along these curves, and Eqs.~\eqref{matrix_elements} and~\eqref{matrix_elements_D} can then be used to calculate the matrix elements in the Fock basis. To show the performance of such a reconstruction, we increase the maximum number $\tilde{N}$ of pulse-sequence segments and reconstruct a density operator $\rho_{n,\tilde{N}}$ from interpolating over all curves $\xi(\tau,N)$ with $N\leq\tilde{N}$. As a measure for the accuracy of $\rho_{n,\tilde{N}}$ we employ its trace distance $d_{n,\tilde{N}}$ to the exact density matrix $\rho_n=|n\rangle\langle n|$ of the Fock state, which is given by $d_{n,\tilde{N}}={\rm Tr}\{\sqrt{(\rho_{n,\tilde{N}}-\rho_n)^2}\}/2$. Figure~\ref{figs4}(b) shows this trace distance for $\tilde{N}=(1,...,20)$, $g/\nu=3/40$, and three values of $n$.

%


\begin{thebibliography}{72}%
\makeatletter
\providecommand \@ifxundefined [1]{%
 \@ifx{#1\undefined}
}%
\providecommand \@ifnum [1]{%
 \ifnum #1\expandafter \@firstoftwo
 \else \expandafter \@secondoftwo
 \fi
}%
\providecommand \@ifx [1]{%
 \ifx #1\expandafter \@firstoftwo
 \else \expandafter \@secondoftwo
 \fi
}%
\providecommand \natexlab [1]{#1}%
\providecommand \enquote  [1]{``#1''}%
\providecommand \bibnamefont  [1]{#1}%
\providecommand \bibfnamefont [1]{#1}%
\providecommand \citenamefont [1]{#1}%
\providecommand \href@noop [0]{\@secondoftwo}%
\providecommand \href [0]{\begingroup \@sanitize@url \@href}%
\providecommand \@href[1]{\@@startlink{#1}\@@href}%
\providecommand \@@href[1]{\endgroup#1\@@endlink}%
\providecommand \@sanitize@url [0]{\catcode `\\12\catcode `\$12\catcode
  `\&12\catcode `\#12\catcode `\^12\catcode `\_12\catcode `\%12\relax}%
\providecommand \@@startlink[1]{}%
\providecommand \@@endlink[0]{}%
\providecommand \url  [0]{\begingroup\@sanitize@url \@url }%
\providecommand \@url [1]{\endgroup\@href {#1}{\urlprefix }}%
\providecommand \urlprefix  [0]{URL }%
\providecommand \Eprint [0]{\href }%
\providecommand \doibase [0]{http://dx.doi.org/}%
\providecommand \selectlanguage [0]{\@gobble}%
\providecommand \bibinfo  [0]{\@secondoftwo}%
\providecommand \bibfield  [0]{\@secondoftwo}%
\providecommand \translation [1]{[#1]}%
\providecommand \BibitemOpen [0]{}%
\providecommand \bibitemStop [0]{}%
\providecommand \bibitemNoStop [0]{.\EOS\space}%
\providecommand \EOS [0]{\spacefactor3000\relax}%
\providecommand \BibitemShut  [1]{\csname bibitem#1\endcsname}%
\let\auto@bib@innerbib\@empty
\bibitem [{\citenamefont {Aspelmeyer}\ \emph {et~al.}(2014)\citenamefont
  {Aspelmeyer}, \citenamefont {Kippenberg},\ and\ \citenamefont
  {Marquardt}}]{Aspelmeyer2014}%
  \BibitemOpen
  \bibfield  {author} {\bibinfo {author} {\bibfnamefont {M.}~\bibnamefont
  {Aspelmeyer}}, \bibinfo {author} {\bibfnamefont {T.~J.}\ \bibnamefont
  {Kippenberg}}, \ and\ \bibinfo {author} {\bibfnamefont {F.}~\bibnamefont
  {Marquardt}},\ }\bibfield  {title} {\enquote {\bibinfo {title} {Cavity
  optomechanics},}\ }\href {\doibase 10.1103/RevModPhys.86.1391} {\bibfield
  {journal} {\bibinfo  {journal} {Rev. Mod. Phys.}\ }\textbf {\bibinfo {volume}
  {86}},\ \bibinfo {pages} {1391} (\bibinfo {year} {2014})}\BibitemShut
  {NoStop}%
\bibitem [{\citenamefont {Neukirch}\ and\ \citenamefont
  {Vamivakas}(2015)}]{Neukirch2015}%
  \BibitemOpen
  \bibfield  {author} {\bibinfo {author} {\bibfnamefont {L.~P.}\ \bibnamefont
  {Neukirch}}\ and\ \bibinfo {author} {\bibfnamefont {A.~N.}\ \bibnamefont
  {Vamivakas}},\ }\bibfield  {title} {\enquote {\bibinfo {title}
  {Nano-optomechanics with optically levitated nanoparticles},}\ }\href
  {\doibase 10.1080/00107514.2014.969492} {\bibfield  {journal} {\bibinfo
  {journal} {Contemp. Phys.}\ }\textbf {\bibinfo {volume} {56}},\ \bibinfo
  {pages} {48} (\bibinfo {year} {2015})}\BibitemShut {NoStop}%
\bibitem [{\citenamefont {Nielsen}\ and\ \citenamefont
  {Chuang}(2003)}]{Nielsen2003}%
  \BibitemOpen
  \bibfield  {author} {\bibinfo {author} {\bibfnamefont {M.~A.}\ \bibnamefont
  {Nielsen}}\ and\ \bibinfo {author} {\bibfnamefont {I.~L.}\ \bibnamefont
  {Chuang}},\ }\href@noop {} {\emph {\bibinfo {title} {Quantum Computation and
  Quantum Information}}}\ (\bibinfo  {publisher} {Cambridge University Press},\
  \bibinfo {address} {Cambridge},\ \bibinfo {year} {2003})\BibitemShut
  {NoStop}%
\bibitem [{\citenamefont {Briegel}\ \emph {et~al.}(2009)\citenamefont
  {Briegel}, \citenamefont {Browne}, \citenamefont {D{\"u}r}, \citenamefont
  {Raussendorf},\ and\ \citenamefont {Van~den Nest}}]{Briegel2009}%
  \BibitemOpen
  \bibfield  {author} {\bibinfo {author} {\bibfnamefont {H.~J.}\ \bibnamefont
  {Briegel}}, \bibinfo {author} {\bibfnamefont {D.~E.}\ \bibnamefont {Browne}},
  \bibinfo {author} {\bibfnamefont {W.}~\bibnamefont {D{\"u}r}}, \bibinfo
  {author} {\bibfnamefont {R.}~\bibnamefont {Raussendorf}}, \ and\ \bibinfo
  {author} {\bibfnamefont {M.}~\bibnamefont {Van~den Nest}},\ }\bibfield
  {title} {\enquote {\bibinfo {title} {Measurement-based quantum
  computation},}\ }\href {\doibase 10.1038/nphys1157} {\bibfield  {journal}
  {\bibinfo  {journal} {Nat. Phys.}\ }\textbf {\bibinfo {volume} {5}},\
  \bibinfo {pages} {19} (\bibinfo {year} {2009})}\BibitemShut {NoStop}%
\bibitem [{\citenamefont {Ladd}\ \emph {et~al.}(2010)\citenamefont {Ladd},
  \citenamefont {Jelezko}, \citenamefont {Laflamme}, \citenamefont {Nakamura},
  \citenamefont {Monroe},\ and\ \citenamefont {O’Brien}}]{Ladd2010}%
  \BibitemOpen
  \bibfield  {author} {\bibinfo {author} {\bibfnamefont {T.~D.}\ \bibnamefont
  {Ladd}}, \bibinfo {author} {\bibfnamefont {F.}~\bibnamefont {Jelezko}},
  \bibinfo {author} {\bibfnamefont {R.}~\bibnamefont {Laflamme}}, \bibinfo
  {author} {\bibfnamefont {Y.}~\bibnamefont {Nakamura}}, \bibinfo {author}
  {\bibfnamefont {C.}~\bibnamefont {Monroe}}, \ and\ \bibinfo {author}
  {\bibfnamefont {J.~L.}\ \bibnamefont {O’Brien}},\ }\bibfield  {title}
  {\enquote {\bibinfo {title} {Quantum computers},}\ }\href {\doibase
  10.1038/nature08812} {\bibfield  {journal} {\bibinfo  {journal} {Nature
  (London)}\ }\textbf {\bibinfo {volume} {464}},\ \bibinfo {pages} {45}
  (\bibinfo {year} {2010})}\BibitemShut {NoStop}%
\bibitem [{\citenamefont {Altepeter}\ \emph {et~al.}(2005)\citenamefont
  {Altepeter}, \citenamefont {Jeffrey},\ and\ \citenamefont
  {Kwiat}}]{Altepeter2005}%
  \BibitemOpen
  \bibfield  {author} {\bibinfo {author} {\bibfnamefont {J.~B.}\ \bibnamefont
  {Altepeter}}, \bibinfo {author} {\bibfnamefont {E.~R.}\ \bibnamefont
  {Jeffrey}}, \ and\ \bibinfo {author} {\bibfnamefont {P.~G.}\ \bibnamefont
  {Kwiat}},\ }\bibfield  {title} {\enquote {\bibinfo {title} {Photonic
  \text{S}tate \text{T}omography},}\ }\href {\doibase
  10.1016/S1049-250X(05)52003-2} {\bibfield  {journal} {\bibinfo  {journal}
  {Adv. At. Mol. Opt. Phys.}\ }\textbf {\bibinfo {volume} {53}},\ \bibinfo
  {pages} {105} (\bibinfo {year} {2005})}\BibitemShut {NoStop}%
\bibitem [{\citenamefont {Del\'eglise}\ \emph {et~al.}(2008)\citenamefont
  {Del\'eglise}, \citenamefont {Dotsenko}, \citenamefont {Sayrin},
  \citenamefont {Bernu}, \citenamefont {Brune}, \citenamefont {Raimond},\ and\
  \citenamefont {Haroche}}]{Deglise2008}%
  \BibitemOpen
  \bibfield  {author} {\bibinfo {author} {\bibfnamefont {S.}~\bibnamefont
  {Del\'eglise}}, \bibinfo {author} {\bibfnamefont {I.}~\bibnamefont
  {Dotsenko}}, \bibinfo {author} {\bibfnamefont {C.}~\bibnamefont {Sayrin}},
  \bibinfo {author} {\bibfnamefont {J.}~\bibnamefont {Bernu}}, \bibinfo
  {author} {\bibfnamefont {M.}~\bibnamefont {Brune}}, \bibinfo {author}
  {\bibfnamefont {J.-M.}\ \bibnamefont {Raimond}}, \ and\ \bibinfo {author}
  {\bibfnamefont {S.}~\bibnamefont {Haroche}},\ }\bibfield  {title} {\enquote
  {\bibinfo {title} {Reconstruction of non-classical cavity field states with
  snapshots of their decoherence},}\ }\href {\doibase 10.1038/nature07288}
  {\bibfield  {journal} {\bibinfo  {journal} {Nature (London)}\ }\textbf
  {\bibinfo {volume} {455}},\ \bibinfo {pages} {510} (\bibinfo {year}
  {2008})}\BibitemShut {NoStop}%
\bibitem [{\citenamefont {Elzerman}\ \emph {et~al.}(2004)\citenamefont
  {Elzerman}, \citenamefont {Hanson}, \citenamefont {Willems~van Beveren},
  \citenamefont {Witkamp}, \citenamefont {Vandersypen},\ and\ \citenamefont
  {Kouwenhoven}}]{Elzerman2004}%
  \BibitemOpen
  \bibfield  {author} {\bibinfo {author} {\bibfnamefont {J.~M.}\ \bibnamefont
  {Elzerman}}, \bibinfo {author} {\bibfnamefont {R.}~\bibnamefont {Hanson}},
  \bibinfo {author} {\bibfnamefont {L.~H.}\ \bibnamefont {Willems~van
  Beveren}}, \bibinfo {author} {\bibfnamefont {B.}~\bibnamefont {Witkamp}},
  \bibinfo {author} {\bibfnamefont {L.~M.~K.}\ \bibnamefont {Vandersypen}}, \
  and\ \bibinfo {author} {\bibfnamefont {L.~P.}\ \bibnamefont {Kouwenhoven}},\
  }\bibfield  {title} {\enquote {\bibinfo {title} {Single-shot read-out of an
  individual electron spin in a quantum dot},}\ }\href {\doibase
  10.1038/nature02693} {\bibfield  {journal} {\bibinfo  {journal} {Nature
  (London)}\ }\textbf {\bibinfo {volume} {430}},\ \bibinfo {pages} {431}
  (\bibinfo {year} {2004})}\BibitemShut {NoStop}%
\bibitem [{\citenamefont {Barthel}\ \emph {et~al.}(2009)\citenamefont
  {Barthel}, \citenamefont {Reilly}, \citenamefont {Marcus}, \citenamefont
  {Hanson},\ and\ \citenamefont {Gossard}}]{Barthel2009}%
  \BibitemOpen
  \bibfield  {author} {\bibinfo {author} {\bibfnamefont {C.}~\bibnamefont
  {Barthel}}, \bibinfo {author} {\bibfnamefont {D.~J.}\ \bibnamefont {Reilly}},
  \bibinfo {author} {\bibfnamefont {C.~M.}\ \bibnamefont {Marcus}}, \bibinfo
  {author} {\bibfnamefont {M.~P.}\ \bibnamefont {Hanson}}, \ and\ \bibinfo
  {author} {\bibfnamefont {A.~C.}\ \bibnamefont {Gossard}},\ }\bibfield
  {title} {\enquote {\bibinfo {title} {Rapid \text{S}ingle-\text{S}hot
  \text{M}easurement of a \text{S}inglet-\text{T}riplet \text{Q}ubit},}\ }\href
  {\doibase 10.1103/PhysRevLett.103.160503} {\bibfield  {journal} {\bibinfo
  {journal} {Phys. Rev. Lett}\ }\textbf {\bibinfo {volume} {103}},\ \bibinfo
  {pages} {160503} (\bibinfo {year} {2009})}\BibitemShut {NoStop}%
\bibitem [{\citenamefont {Morello}\ \emph {et~al.}(2010)\citenamefont
  {Morello}, \citenamefont {Pla}, \citenamefont {Zwanenburg}, \citenamefont
  {Chan}, \citenamefont {Tan}, \citenamefont {Huebl}, \citenamefont
  {{M{\"o}tt{\"o}nen}}, \citenamefont {Nugroho}, \citenamefont {Yang},
  \citenamefont {van Donkelaar}, \citenamefont {Alves}, \citenamefont
  {Jamieson}, \citenamefont {Escott}, \citenamefont {Hollenberg}, \citenamefont
  {Clark},\ and\ \citenamefont {Dzurak}}]{Morello2010}%
  \BibitemOpen
  \bibfield  {author} {\bibinfo {author} {\bibfnamefont {A.}~\bibnamefont
  {Morello}}, \bibinfo {author} {\bibfnamefont {J.~J.}\ \bibnamefont {Pla}},
  \bibinfo {author} {\bibfnamefont {F.~A.}\ \bibnamefont {Zwanenburg}},
  \bibinfo {author} {\bibfnamefont {K.~W.}\ \bibnamefont {Chan}}, \bibinfo
  {author} {\bibfnamefont {K.~Y.}\ \bibnamefont {Tan}}, \bibinfo {author}
  {\bibfnamefont {H.}~\bibnamefont {Huebl}}, \bibinfo {author} {\bibfnamefont
  {M.}~\bibnamefont {{M{\"o}tt{\"o}nen}}}, \bibinfo {author} {\bibfnamefont
  {C.~D.}\ \bibnamefont {Nugroho}}, \bibinfo {author} {\bibfnamefont
  {C.}~\bibnamefont {Yang}}, \bibinfo {author} {\bibfnamefont {J.~A.}\
  \bibnamefont {van Donkelaar}}, \bibinfo {author} {\bibfnamefont {A.~D.~C.}\
  \bibnamefont {Alves}}, \bibinfo {author} {\bibfnamefont {D.~N.}\ \bibnamefont
  {Jamieson}}, \bibinfo {author} {\bibfnamefont {C.~C.}\ \bibnamefont
  {Escott}}, \bibinfo {author} {\bibfnamefont {L.~C.~L.}\ \bibnamefont
  {Hollenberg}}, \bibinfo {author} {\bibfnamefont {R.~G.}\ \bibnamefont
  {Clark}}, \ and\ \bibinfo {author} {\bibfnamefont {A.~S.}\ \bibnamefont
  {Dzurak}},\ }\bibfield  {title} {\enquote {\bibinfo {title} {Single-shot
  readout of an electron spin in silicon},}\ }\href {\doibase
  10.1038/nature09392} {\bibfield  {journal} {\bibinfo  {journal} {Nature
  (London)}\ }\textbf {\bibinfo {volume} {467}},\ \bibinfo {pages} {687}
  (\bibinfo {year} {2010})}\BibitemShut {NoStop}%
\bibitem [{\citenamefont {Bochmann}\ \emph {et~al.}(2010)\citenamefont
  {Bochmann}, \citenamefont {M{\"u}cke}, \citenamefont {Guhl}, \citenamefont
  {Ritter}, \citenamefont {Rempe},\ and\ \citenamefont
  {Moehring}}]{Bochmann2010}%
  \BibitemOpen
  \bibfield  {author} {\bibinfo {author} {\bibfnamefont {J.}~\bibnamefont
  {Bochmann}}, \bibinfo {author} {\bibfnamefont {M.}~\bibnamefont {M{\"u}cke}},
  \bibinfo {author} {\bibfnamefont {C.}~\bibnamefont {Guhl}}, \bibinfo {author}
  {\bibfnamefont {S.}~\bibnamefont {Ritter}}, \bibinfo {author} {\bibfnamefont
  {G.}~\bibnamefont {Rempe}}, \ and\ \bibinfo {author} {\bibfnamefont {D.~L.}\
  \bibnamefont {Moehring}},\ }\bibfield  {title} {\enquote {\bibinfo {title}
  {Lossless \text{S}tate \text{D}etection of \text{S}ingle \text{N}eutral
  \text{A}toms},}\ }\href {\doibase 10.1103/PhysRevLett.104.203601} {\bibfield
  {journal} {\bibinfo  {journal} {Phys. Rev. Lett}\ }\textbf {\bibinfo {volume}
  {104}},\ \bibinfo {pages} {203601} (\bibinfo {year} {2010})}\BibitemShut
  {NoStop}%
\bibitem [{\citenamefont {Steffen}\ \emph {et~al.}(2006)\citenamefont
  {Steffen}, \citenamefont {Ansmann}, \citenamefont {Bialczak}, \citenamefont
  {Katz}, \citenamefont {Lucero}, \citenamefont {Mcdermott}, \citenamefont
  {Neeley}, \citenamefont {Weig}, \citenamefont {Cleland},\ and\ \citenamefont
  {Martinis}}]{Steffen2006}%
  \BibitemOpen
  \bibfield  {author} {\bibinfo {author} {\bibfnamefont {M.}~\bibnamefont
  {Steffen}}, \bibinfo {author} {\bibfnamefont {M.}~\bibnamefont {Ansmann}},
  \bibinfo {author} {\bibfnamefont {R.~C.}\ \bibnamefont {Bialczak}}, \bibinfo
  {author} {\bibfnamefont {N.}~\bibnamefont {Katz}}, \bibinfo {author}
  {\bibfnamefont {E.}~\bibnamefont {Lucero}}, \bibinfo {author} {\bibfnamefont
  {R.}~\bibnamefont {Mcdermott}}, \bibinfo {author} {\bibfnamefont
  {M.}~\bibnamefont {Neeley}}, \bibinfo {author} {\bibfnamefont {E.~M.}\
  \bibnamefont {Weig}}, \bibinfo {author} {\bibfnamefont {A.~N.}\ \bibnamefont
  {Cleland}}, \ and\ \bibinfo {author} {\bibfnamefont {J.~M.}\ \bibnamefont
  {Martinis}},\ }\bibfield  {title} {\enquote {\bibinfo {title} {Measurement of
  the \text{E}ntanglement of \text{T}wo \text{S}uperconducting \text{Q}ubits
  \text{V}ia \text{S}tate \text{T}omography},}\ }\href {\doibase
  10.1126/science.1130886} {\bibfield  {journal} {\bibinfo  {journal}
  {Science}\ }\textbf {\bibinfo {volume} {313}},\ \bibinfo {pages} {1423}
  (\bibinfo {year} {2006})}\BibitemShut {NoStop}%
\bibitem [{\citenamefont {Fedorov}\ \emph {et~al.}(2014)\citenamefont
  {Fedorov}, \citenamefont {Shcherbakova}, \citenamefont {Wolf}, \citenamefont
  {Beckmann},\ and\ \citenamefont {Ustinov}}]{Fedorov2010}%
  \BibitemOpen
  \bibfield  {author} {\bibinfo {author} {\bibfnamefont {K.~G.}\ \bibnamefont
  {Fedorov}}, \bibinfo {author} {\bibfnamefont {A.~V.}\ \bibnamefont
  {Shcherbakova}}, \bibinfo {author} {\bibfnamefont {M.~J.}\ \bibnamefont
  {Wolf}}, \bibinfo {author} {\bibfnamefont {D.}~\bibnamefont {Beckmann}}, \
  and\ \bibinfo {author} {\bibfnamefont {A.~V.}\ \bibnamefont {Ustinov}},\
  }\bibfield  {title} {\enquote {\bibinfo {title} {Fluxon \text{R}eadout of a
  \text{S}uperconducting \text{Q}ubit},}\ }\href {\doibase
  10.1103/PhysRevLett.112.160502} {\bibfield  {journal} {\bibinfo  {journal}
  {Phys. Rev. Lett}\ }\textbf {\bibinfo {volume} {112}},\ \bibinfo {pages}
  {160502} (\bibinfo {year} {2014})}\BibitemShut {NoStop}%
\bibitem [{\citenamefont {Vanner}\ \emph {et~al.}(2013)\citenamefont {Vanner},
  \citenamefont {Hofer}, \citenamefont {Cole},\ and\ \citenamefont
  {Aspelmeyer}}]{Vanner2013}%
  \BibitemOpen
  \bibfield  {author} {\bibinfo {author} {\bibfnamefont {M.~R.}\ \bibnamefont
  {Vanner}}, \bibinfo {author} {\bibfnamefont {J.}~\bibnamefont {Hofer}},
  \bibinfo {author} {\bibfnamefont {G.~D.}\ \bibnamefont {Cole}}, \ and\
  \bibinfo {author} {\bibfnamefont {M.}~\bibnamefont {Aspelmeyer}},\ }\bibfield
   {title} {\enquote {\bibinfo {title} {Cooling-by-measurement and mechanical
  state tomography via pulsed optomechanics},}\ }\href {\doibase
  10.1038/ncomms3295} {\bibfield  {journal} {\bibinfo  {journal} {Nat.
  Commun.}\ }\textbf {\bibinfo {volume} {4}},\ \bibinfo {pages} {2295}
  (\bibinfo {year} {2013})}\BibitemShut {NoStop}%
\bibitem [{\citenamefont {Rashid}\ \emph {et~al.}(2017)\citenamefont {Rashid},
  \citenamefont {Toro\v{s}},\ and\ \citenamefont {Ulbricht}}]{Rashid2017}%
  \BibitemOpen
  \bibfield  {author} {\bibinfo {author} {\bibfnamefont {M.}~\bibnamefont
  {Rashid}}, \bibinfo {author} {\bibfnamefont {M.}~\bibnamefont {Toro\v{s}}}, \
  and\ \bibinfo {author} {\bibfnamefont {H.}~\bibnamefont {Ulbricht}},\
  }\bibfield  {title} {\enquote {\bibinfo {title} {Wigner function
  reconstruction in levitated optomechanics},}\ }\href {\doibase
  10.1515/qmetro-2017-0003} {\bibfield  {journal} {\bibinfo  {journal} {Quantum
  Meas. Quantum Metrol.}\ }\textbf {\bibinfo {volume} {4}},\ \bibinfo {pages}
  {17} (\bibinfo {year} {2017})}\BibitemShut {NoStop}%
\bibitem [{\citenamefont {Leibfried}\ \emph {et~al.}(1996)\citenamefont
  {Leibfried}, \citenamefont {Meekhof}, \citenamefont {King}, \citenamefont
  {Monroe}, \citenamefont {Itano},\ and\ \citenamefont
  {Wineland}}]{Leibfried1996}%
  \BibitemOpen
  \bibfield  {author} {\bibinfo {author} {\bibfnamefont {D.}~\bibnamefont
  {Leibfried}}, \bibinfo {author} {\bibfnamefont {D.~M.}\ \bibnamefont
  {Meekhof}}, \bibinfo {author} {\bibfnamefont {B.~E.}\ \bibnamefont {King}},
  \bibinfo {author} {\bibfnamefont {C.}~\bibnamefont {Monroe}}, \bibinfo
  {author} {\bibfnamefont {W.~M.}\ \bibnamefont {Itano}}, \ and\ \bibinfo
  {author} {\bibfnamefont {D.~J.}\ \bibnamefont {Wineland}},\ }\bibfield
  {title} {\enquote {\bibinfo {title} {Experimental \text{D}etermination of the
  \text{M}otional \text{Q}uantum \text{S}tate of a \text{T}rapped
  \text{A}tom},}\ }\href {\doibase 10.1103/PhysRevLett.77.4281} {\bibfield
  {journal} {\bibinfo  {journal} {Phys. Rev. Lett.}\ }\textbf {\bibinfo
  {volume} {77}},\ \bibinfo {pages} {4281} (\bibinfo {year}
  {1996})}\BibitemShut {NoStop}%
\bibitem [{\citenamefont {Bertet}\ \emph {et~al.}(2002)\citenamefont {Bertet},
  \citenamefont {Auff\`eves}, \citenamefont {Maioli}, \citenamefont {Osnaghi},
  \citenamefont {Meunier}, \citenamefont {Brune}, \citenamefont {Raimond},\
  and\ \citenamefont {Haroche}}]{Bertet2002}%
  \BibitemOpen
  \bibfield  {author} {\bibinfo {author} {\bibfnamefont {P.}~\bibnamefont
  {Bertet}}, \bibinfo {author} {\bibfnamefont {A.}~\bibnamefont {Auff\`eves}},
  \bibinfo {author} {\bibfnamefont {P.}~\bibnamefont {Maioli}}, \bibinfo
  {author} {\bibfnamefont {S.}~\bibnamefont {Osnaghi}}, \bibinfo {author}
  {\bibfnamefont {T.}~\bibnamefont {Meunier}}, \bibinfo {author} {\bibfnamefont
  {M.}~\bibnamefont {Brune}}, \bibinfo {author} {\bibfnamefont {J.~M.}\
  \bibnamefont {Raimond}}, \ and\ \bibinfo {author} {\bibfnamefont
  {S.}~\bibnamefont {Haroche}},\ }\bibfield  {title} {\enquote {\bibinfo
  {title} {Direct \text{M}easurement of the \text{W}igner \text{F}unction of a
  \text{O}ne-\text{P}hoton \text{F}ock \text{S}tate in a \text{C}avity},}\
  }\href {\doibase 10.1103/PhysRevLett.89.200402} {\bibfield  {journal}
  {\bibinfo  {journal} {Phys. Rev. Lett.}\ }\textbf {\bibinfo {volume} {89}},\
  \bibinfo {pages} {200402} (\bibinfo {year} {2002})}\BibitemShut {NoStop}%
\bibitem [{\citenamefont {Wallraff}\ \emph {et~al.}(2005)\citenamefont
  {Wallraff}, \citenamefont {Schuster}, \citenamefont {Blais}, \citenamefont
  {Frunzio}, \citenamefont {Majer}, \citenamefont {Devoret}, \citenamefont
  {Girvin},\ and\ \citenamefont {Schoelkopf}}]{Wallraff2005}%
  \BibitemOpen
  \bibfield  {author} {\bibinfo {author} {\bibfnamefont {A.}~\bibnamefont
  {Wallraff}}, \bibinfo {author} {\bibfnamefont {D.~I.}\ \bibnamefont
  {Schuster}}, \bibinfo {author} {\bibfnamefont {A.}~\bibnamefont {Blais}},
  \bibinfo {author} {\bibfnamefont {L.}~\bibnamefont {Frunzio}}, \bibinfo
  {author} {\bibfnamefont {J.}~\bibnamefont {Majer}}, \bibinfo {author}
  {\bibfnamefont {M.~H.}\ \bibnamefont {Devoret}}, \bibinfo {author}
  {\bibfnamefont {S.M.}\ \bibnamefont {Girvin}}, \ and\ \bibinfo {author}
  {\bibfnamefont {R.~J.}\ \bibnamefont {Schoelkopf}},\ }\bibfield  {title}
  {\enquote {\bibinfo {title} {Approaching \text{U}nit \text{V}isibility for
  \text{C}ontrol of a \text{S}uperconducting \text{Q}ubit with
  \text{D}ispersive \text{R}eadout},}\ }\href {\doibase
  10.1103/PhysRevLett.95.060501} {\bibfield  {journal} {\bibinfo  {journal}
  {Phys. Rev. Lett}\ }\textbf {\bibinfo {volume} {95}},\ \bibinfo {pages}
  {060501} (\bibinfo {year} {2005})}\BibitemShut {NoStop}%
\bibitem [{\citenamefont {Burgarth}\ \emph {et~al.}(2011)\citenamefont
  {Burgarth}, \citenamefont {Maruyama},\ and\ \citenamefont
  {Nori}}]{Burgarth2011}%
  \BibitemOpen
  \bibfield  {author} {\bibinfo {author} {\bibfnamefont {D.}~\bibnamefont
  {Burgarth}}, \bibinfo {author} {\bibfnamefont {K.}~\bibnamefont {Maruyama}},
  \ and\ \bibinfo {author} {\bibfnamefont {F.}~\bibnamefont {Nori}},\
  }\bibfield  {title} {\enquote {\bibinfo {title} {Indirect quantum tomography
  of quadratic \text{H}amiltonians},}\ }\href {\doibase
  10.1088/1367-2630/13/1/013019} {\bibfield  {journal} {\bibinfo  {journal}
  {New J. Phys.}\ }\textbf {\bibinfo {volume} {13}},\ \bibinfo {pages} {013019}
  (\bibinfo {year} {2011})}\BibitemShut {NoStop}%
\bibitem [{\citenamefont {Burgarth}\ \emph {et~al.}(2015)\citenamefont
  {Burgarth}, \citenamefont {Giovannetti}, \citenamefont {Kato},\ and\
  \citenamefont {Yuasa}}]{Burgarth2015}%
  \BibitemOpen
  \bibfield  {author} {\bibinfo {author} {\bibfnamefont {D.}~\bibnamefont
  {Burgarth}}, \bibinfo {author} {\bibfnamefont {V.}~\bibnamefont
  {Giovannetti}}, \bibinfo {author} {\bibfnamefont {A.~N.}\ \bibnamefont
  {Kato}}, \ and\ \bibinfo {author} {\bibfnamefont {K.}~\bibnamefont {Yuasa}},\
  }\bibfield  {title} {\enquote {\bibinfo {title} {Quantum estimation via
  sequential measurements},}\ }\href {\doibase 10.1088/1367-2630/17/11/113055}
  {\bibfield  {journal} {\bibinfo  {journal} {New J. Phys.}\ }\textbf {\bibinfo
  {volume} {17}},\ \bibinfo {pages} {113055} (\bibinfo {year}
  {2015})}\BibitemShut {NoStop}%
\bibitem [{\citenamefont {Sone}\ and\ \citenamefont
  {Cappellaro}(2017{\natexlab{a}})}]{Sone2017a}%
  \BibitemOpen
  \bibfield  {author} {\bibinfo {author} {\bibfnamefont {A.}~\bibnamefont
  {Sone}}\ and\ \bibinfo {author} {\bibfnamefont {P.}~\bibnamefont
  {Cappellaro}},\ }\bibfield  {title} {\enquote {\bibinfo {title} {Exact
  dimension estimation of interacting qubit systems assisted by a single
  quantum probe},}\ }\href {\doibase 10.1103/PhysRevA.96.062334} {\bibfield
  {journal} {\bibinfo  {journal} {Phys. Rev. A}\ }\textbf {\bibinfo {volume}
  {96}},\ \bibinfo {pages} {062334} (\bibinfo {year}
  {2017}{\natexlab{a}})}\BibitemShut {NoStop}%
\bibitem [{\citenamefont {Sone}\ and\ \citenamefont
  {Cappellaro}(2017{\natexlab{b}})}]{Sone2017b}%
  \BibitemOpen
  \bibfield  {author} {\bibinfo {author} {\bibfnamefont {A.}~\bibnamefont
  {Sone}}\ and\ \bibinfo {author} {\bibfnamefont {P.}~\bibnamefont
  {Cappellaro}},\ }\bibfield  {title} {\enquote {\bibinfo {title} {Hamiltonian
  identifiability assisted by a single-probe measurement},}\ }\href {\doibase
  10.1103/PhysRevA.95.022335} {\bibfield  {journal} {\bibinfo  {journal} {Phys.
  Rev. A}\ }\textbf {\bibinfo {volume} {95}},\ \bibinfo {pages} {022335}
  (\bibinfo {year} {2017}{\natexlab{b}})}\BibitemShut {NoStop}%
\bibitem [{\citenamefont {Zhang}\ \emph {et~al.}()\citenamefont {Zhang},
  \citenamefont {Hegde},\ and\ \citenamefont {Suter}}]{Zhang2018}%
  \BibitemOpen
  \bibfield  {author} {\bibinfo {author} {\bibfnamefont {J.}~\bibnamefont
  {Zhang}}, \bibinfo {author} {\bibfnamefont {S.~S.}\ \bibnamefont {Hegde}}, \
  and\ \bibinfo {author} {\bibfnamefont {D.}~\bibnamefont {Suter}},\ }\bibfield
   {title} {\enquote {\bibinfo {title} {Efficient indirect control of nuclear
  spins in diamond \text{NV} centers},}\ }\href
  {https://arxiv.org/abs/1810.08230} {\bibinfo  {journal} {arXiv:1810.08230}\
  }\BibitemShut {NoStop}%
\bibitem [{\citenamefont {Cronin}\ \emph {et~al.}(2009)\citenamefont {Cronin},
  \citenamefont {Schmiedmayer},\ and\ \citenamefont {Pritchard}}]{Cronin2009}%
  \BibitemOpen
\bibfield  {journal} {  }\bibfield  {author} {\bibinfo {author} {\bibfnamefont
  {A.~D.}\ \bibnamefont {Cronin}}, \bibinfo {author} {\bibfnamefont
  {J.}~\bibnamefont {Schmiedmayer}}, \ and\ \bibinfo {author} {\bibfnamefont
  {D.~E.}\ \bibnamefont {Pritchard}},\ }\bibfield  {title} {\enquote {\bibinfo
  {title} {Optics and interferometry with atoms and molecules},}\ }\href
  {\doibase 10.1103/RevModPhys.81.1051} {\bibfield  {journal} {\bibinfo
  {journal} {Rev. Mod. Phys.}\ }\textbf {\bibinfo {volume} {81}},\ \bibinfo
  {pages} {1051} (\bibinfo {year} {2009})}\BibitemShut {NoStop}%
\bibitem [{\citenamefont {Recati}\ \emph {et~al.}(2005)\citenamefont {Recati},
  \citenamefont {Fedichev}, \citenamefont {Zwerger}, \citenamefont {von
  Delft},\ and\ \citenamefont {Zoller}}]{Recati2005}%
  \BibitemOpen
  \bibfield  {author} {\bibinfo {author} {\bibfnamefont {A.}~\bibnamefont
  {Recati}}, \bibinfo {author} {\bibfnamefont {P.~O.}\ \bibnamefont
  {Fedichev}}, \bibinfo {author} {\bibfnamefont {W.}~\bibnamefont {Zwerger}},
  \bibinfo {author} {\bibfnamefont {J.}~\bibnamefont {von Delft}}, \ and\
  \bibinfo {author} {\bibfnamefont {P.}~\bibnamefont {Zoller}},\ }\bibfield
  {title} {\enquote {\bibinfo {title} {Atomic \text{Q}uantum \text{D}ots
  \text{C}oupled to a \text{R}eservoir of a \text{S}uperfluid
  \text{B}ose-\text{E}instein \text{C}ondensate},}\ }\href {\doibase
  10.1103/PhysRevLett.94.040404} {\bibfield  {journal} {\bibinfo  {journal}
  {Phys. Rev. Lett.}\ }\textbf {\bibinfo {volume} {94}},\ \bibinfo {pages}
  {040404} (\bibinfo {year} {2005})}\BibitemShut {NoStop}%
\bibitem [{\citenamefont {Quan}\ \emph {et~al.}(2006)\citenamefont {Quan},
  \citenamefont {Song}, \citenamefont {Liu}, \citenamefont {Zanardi},\ and\
  \citenamefont {Sun}}]{Quan2006}%
  \BibitemOpen
  \bibfield  {author} {\bibinfo {author} {\bibfnamefont {H.~T.}\ \bibnamefont
  {Quan}}, \bibinfo {author} {\bibfnamefont {Z.}~\bibnamefont {Song}}, \bibinfo
  {author} {\bibfnamefont {X.~F.}\ \bibnamefont {Liu}}, \bibinfo {author}
  {\bibfnamefont {P.}~\bibnamefont {Zanardi}}, \ and\ \bibinfo {author}
  {\bibfnamefont {C.~P.}\ \bibnamefont {Sun}},\ }\bibfield  {title} {\enquote
  {\bibinfo {title} {Decay of \text{L}oschmidt \text{E}cho \text{E}nhanced by
  \text{Q}uantum \text{C}riticality},}\ }\href {\doibase
  10.1103/PhysRevLett.96.140604} {\bibfield  {journal} {\bibinfo  {journal}
  {Phys. Rev. Lett.}\ }\textbf {\bibinfo {volume} {96}},\ \bibinfo {pages}
  {140604} (\bibinfo {year} {2006})}\BibitemShut {NoStop}%
\bibitem [{\citenamefont {Dorner}\ \emph {et~al.}(2013)\citenamefont {Dorner},
  \citenamefont {Clark}, \citenamefont {Heaney}, \citenamefont {Fazio},
  \citenamefont {Goold},\ and\ \citenamefont {Vedral}}]{Dorner2013}%
  \BibitemOpen
  \bibfield  {author} {\bibinfo {author} {\bibfnamefont {R.}~\bibnamefont
  {Dorner}}, \bibinfo {author} {\bibfnamefont {S.~R.}\ \bibnamefont {Clark}},
  \bibinfo {author} {\bibfnamefont {L.}~\bibnamefont {Heaney}}, \bibinfo
  {author} {\bibfnamefont {R.}~\bibnamefont {Fazio}}, \bibinfo {author}
  {\bibfnamefont {J.}~\bibnamefont {Goold}}, \ and\ \bibinfo {author}
  {\bibfnamefont {V.}~\bibnamefont {Vedral}},\ }\bibfield  {title} {\enquote
  {\bibinfo {title} {Extracting \text{Q}uantum \text{W}ork \text{S}tatistics
  and \text{F}luctuation \text{T}heorems by \text{S}ingle-\text{Q}ubit
  \text{I}nterferometry},}\ }\href {\doibase 10.1103/PhysRevLett.110.230601}
  {\bibfield  {journal} {\bibinfo  {journal} {Phys. Rev. Lett.}\ }\textbf
  {\bibinfo {volume} {110}},\ \bibinfo {pages} {230601} (\bibinfo {year}
  {2013})}\BibitemShut {NoStop}%
\bibitem [{\citenamefont {Peng}\ \emph {et~al.}(2015)\citenamefont {Peng},
  \citenamefont {Zhou}, \citenamefont {Wei}, \citenamefont {Cui}, \citenamefont
  {Du},\ and\ \citenamefont {Liu}}]{Peng2015}%
  \BibitemOpen
  \bibfield  {author} {\bibinfo {author} {\bibfnamefont {X.}~\bibnamefont
  {Peng}}, \bibinfo {author} {\bibfnamefont {H.}~\bibnamefont {Zhou}}, \bibinfo
  {author} {\bibfnamefont {B.-B.}\ \bibnamefont {Wei}}, \bibinfo {author}
  {\bibfnamefont {J.}~\bibnamefont {Cui}}, \bibinfo {author} {\bibfnamefont
  {J.}~\bibnamefont {Du}}, \ and\ \bibinfo {author} {\bibfnamefont {R.-B.}\
  \bibnamefont {Liu}},\ }\bibfield  {title} {\enquote {\bibinfo {title}
  {Experimental \text{O}bservation of \text{L}ee-\text{Y}ang \text{Z}eros},}\
  }\href {\doibase 10.1103/PhysRevLett.114.010601} {\bibfield  {journal}
  {\bibinfo  {journal} {Phys. Rev. Lett.}\ }\textbf {\bibinfo {volume} {114}},\
  \bibinfo {pages} {010601} (\bibinfo {year} {2015})}\BibitemShut {NoStop}%
\bibitem [{\citenamefont {Correa}\ \emph {et~al.}(2015)\citenamefont {Correa},
  \citenamefont {Mehboudi}, \citenamefont {Adesso},\ and\ \citenamefont
  {Sanpera}}]{Correa2015}%
  \BibitemOpen
  \bibfield  {author} {\bibinfo {author} {\bibfnamefont {L.~A.}\ \bibnamefont
  {Correa}}, \bibinfo {author} {\bibfnamefont {M.}~\bibnamefont {Mehboudi}},
  \bibinfo {author} {\bibfnamefont {G.}~\bibnamefont {Adesso}}, \ and\ \bibinfo
  {author} {\bibfnamefont {A.}~\bibnamefont {Sanpera}},\ }\bibfield  {title}
  {\enquote {\bibinfo {title} {Individual \text{Q}uantum \text{P}robes for
  \text{O}ptimal \text{T}hermometry},}\ }\href {\doibase
  10.1103/PhysRevLett.114.220405} {\bibfield  {journal} {\bibinfo  {journal}
  {Phys. Rev. Lett.}\ }\textbf {\bibinfo {volume} {114}},\ \bibinfo {pages}
  {220405} (\bibinfo {year} {2015})}\BibitemShut {NoStop}%
\bibitem [{\citenamefont {Asadian}\ \emph {et~al.}(2014)\citenamefont
  {Asadian}, \citenamefont {Brukner},\ and\ \citenamefont
  {Rabl}}]{Asadian2014}%
  \BibitemOpen
  \bibfield  {author} {\bibinfo {author} {\bibfnamefont {A.}~\bibnamefont
  {Asadian}}, \bibinfo {author} {\bibfnamefont {C.}~\bibnamefont {Brukner}}, \
  and\ \bibinfo {author} {\bibfnamefont {P.}~\bibnamefont {Rabl}},\ }\bibfield
  {title} {\enquote {\bibinfo {title} {Probing \text{M}acroscopic
  \text{R}ealism \text{V}ia \text{R}amsey \text{C}orrelation
  \text{M}easurements},}\ }\href {\doibase 10.1103/PhysRevLett.112.190402}
  {\bibfield  {journal} {\bibinfo  {journal} {Phys. Rev. Lett.}\ }\textbf
  {\bibinfo {volume} {112}},\ \bibinfo {pages} {190402} (\bibinfo {year}
  {2014})}\BibitemShut {NoStop}%
\bibitem [{\citenamefont {Lutterbach}\ and\ \citenamefont
  {Davidovich}(1997)}]{Lutterbach1997}%
  \BibitemOpen
  \bibfield  {author} {\bibinfo {author} {\bibfnamefont {L.~G.}\ \bibnamefont
  {Lutterbach}}\ and\ \bibinfo {author} {\bibfnamefont {L.}~\bibnamefont
  {Davidovich}},\ }\bibfield  {title} {\enquote {\bibinfo {title} {Method for
  \text{D}irect \text{M}easurement of the \text{W}igner \text{F}unction in
  \text{C}avity \text{QED} and \text{I}on \text{T}raps},}\ }\href {\doibase
  10.1103/PhysRevLett.78.2547} {\bibfield  {journal} {\bibinfo  {journal}
  {Phys. Rev. Lett.}\ }\textbf {\bibinfo {volume} {78}},\ \bibinfo {pages}
  {2547} (\bibinfo {year} {1997})}\BibitemShut {NoStop}%
\bibitem [{\citenamefont {Kim}\ \emph {et~al.}(1998)\citenamefont {Kim},
  \citenamefont {Antesberger}, \citenamefont {Bodendorf},\ and\ \citenamefont
  {Walther}}]{Kim1998}%
  \BibitemOpen
  \bibfield  {author} {\bibinfo {author} {\bibfnamefont {M.~S.}\ \bibnamefont
  {Kim}}, \bibinfo {author} {\bibfnamefont {G.}~\bibnamefont {Antesberger}},
  \bibinfo {author} {\bibfnamefont {C.~T.}\ \bibnamefont {Bodendorf}}, \ and\
  \bibinfo {author} {\bibfnamefont {H.}~\bibnamefont {Walther}},\ }\bibfield
  {title} {\enquote {\bibinfo {title} {Scheme for direct observation of the
  \text{W}igner characteristic function in cavity \text{QED}},}\ }\href
  {\doibase 10.1103/PhysRevA.58.R65} {\bibfield  {journal} {\bibinfo  {journal}
  {Phys. Rev. A}\ }\textbf {\bibinfo {volume} {58}},\ \bibinfo {pages} {R65}
  (\bibinfo {year} {1998})}\BibitemShut {NoStop}%
\bibitem [{\citenamefont {Singh}\ and\ \citenamefont
  {Meystre}(2010)}]{Singh2010}%
  \BibitemOpen
  \bibfield  {author} {\bibinfo {author} {\bibfnamefont {S.}~\bibnamefont
  {Singh}}\ and\ \bibinfo {author} {\bibfnamefont {P.}~\bibnamefont
  {Meystre}},\ }\bibfield  {title} {\enquote {\bibinfo {title} {Atomic probe
  \text{W}igner tomography of a nanomechanical system},}\ }\href {\doibase
  10.1103/PhysRevA.81.041804} {\bibfield  {journal} {\bibinfo  {journal} {Phys.
  Rev. A}\ }\textbf {\bibinfo {volume} {81}},\ \bibinfo {pages} {041804}
  (\bibinfo {year} {2010})}\BibitemShut {NoStop}%
\bibitem [{\citenamefont {Casanova}\ \emph {et~al.}(2012)\citenamefont
  {Casanova}, \citenamefont {L{\'o}pez}, \citenamefont {Garc{\'i}a-Ripoll},
  \citenamefont {Roos},\ and\ \citenamefont {Solano}}]{Casanova2012}%
  \BibitemOpen
  \bibfield  {author} {\bibinfo {author} {\bibfnamefont {J.}~\bibnamefont
  {Casanova}}, \bibinfo {author} {\bibfnamefont {C.~E.}\ \bibnamefont
  {L{\'o}pez}}, \bibinfo {author} {\bibfnamefont {J.~J.}\ \bibnamefont
  {Garc{\'i}a-Ripoll}}, \bibinfo {author} {\bibfnamefont {C.~F.}\ \bibnamefont
  {Roos}}, \ and\ \bibinfo {author} {\bibfnamefont {E.}~\bibnamefont
  {Solano}},\ }\bibfield  {title} {\enquote {\bibinfo {title} {Quantum
  tomography in position and momentum space},}\ }\href {\doibase
  10.1140/epjd/e2012-30016-6} {\bibfield  {journal} {\bibinfo  {journal} {Eur.
  Phys. J. D}\ }\textbf {\bibinfo {volume} {66}},\ \bibinfo {pages} {222}
  (\bibinfo {year} {2012})}\BibitemShut {NoStop}%
\bibitem [{\citenamefont {Mazzola}\ \emph {et~al.}(2013)\citenamefont
  {Mazzola}, \citenamefont {De~Chiara},\ and\ \citenamefont
  {Paternostro}}]{Mazzola2013}%
  \BibitemOpen
  \bibfield  {author} {\bibinfo {author} {\bibfnamefont {L.}~\bibnamefont
  {Mazzola}}, \bibinfo {author} {\bibfnamefont {G.}~\bibnamefont {De~Chiara}},
  \ and\ \bibinfo {author} {\bibfnamefont {M.}~\bibnamefont {Paternostro}},\
  }\bibfield  {title} {\enquote {\bibinfo {title} {Measuring the
  \text{C}haracteristic \text{F}unction of the \text{W}ork
  \text{D}istribution},}\ }\href {\doibase 10.1103/PhysRevLett.110.230602}
  {\bibfield  {journal} {\bibinfo  {journal} {Phys. Rev. Lett.}\ }\textbf
  {\bibinfo {volume} {110}},\ \bibinfo {pages} {230602} (\bibinfo {year}
  {2013})}\BibitemShut {NoStop}%
\bibitem [{\citenamefont {Taketani}\ \emph {et~al.}(2014)\citenamefont
  {Taketani}, \citenamefont {Fogarty}, \citenamefont {Kajari}, \citenamefont
  {Busch},\ and\ \citenamefont {Morigi}}]{Taketani2014}%
  \BibitemOpen
  \bibfield  {author} {\bibinfo {author} {\bibfnamefont {B.~G.}\ \bibnamefont
  {Taketani}}, \bibinfo {author} {\bibfnamefont {T.}~\bibnamefont {Fogarty}},
  \bibinfo {author} {\bibfnamefont {E.}~\bibnamefont {Kajari}}, \bibinfo
  {author} {\bibfnamefont {Th.}\ \bibnamefont {Busch}}, \ and\ \bibinfo
  {author} {\bibfnamefont {G.}~\bibnamefont {Morigi}},\ }\bibfield  {title}
  {\enquote {\bibinfo {title} {Quantum reservoirs with ion chains},}\ }\href
  {\doibase 10.1103/PhysRevA.90.012312} {\bibfield  {journal} {\bibinfo
  {journal} {Phys. Rev. A}\ }\textbf {\bibinfo {volume} {90}},\ \bibinfo
  {pages} {012312} (\bibinfo {year} {2014})}\BibitemShut {NoStop}%
\bibitem [{\citenamefont {Uhrig}(2007)}]{Uhri2007}%
  \BibitemOpen
  \bibfield  {author} {\bibinfo {author} {\bibfnamefont {G.~S.}\ \bibnamefont
  {Uhrig}},\ }\bibfield  {title} {\enquote {\bibinfo {title} {Keeping a
  \text{Q}uantum \text{B}it \text{A}live by \text{O}ptimized $\pi$-\text{P}ulse
  \text{S}equences},}\ }\href {\doibase 10.1103/PhysRevLett.98.100504}
  {\bibfield  {journal} {\bibinfo  {journal} {Phys. Rev. Lett.}\ }\textbf
  {\bibinfo {volume} {98}},\ \bibinfo {pages} {100504} (\bibinfo {year}
  {2007})}\BibitemShut {NoStop}%
\bibitem [{\citenamefont {de~Lange}\ \emph {et~al.}(2010)\citenamefont
  {de~Lange}, \citenamefont {Wang}, \citenamefont {Rist{\`e}}, \citenamefont
  {Dobrovitski},\ and\ \citenamefont {Hanson}}]{de-Lange2010}%
  \BibitemOpen
  \bibfield  {author} {\bibinfo {author} {\bibfnamefont {G.}~\bibnamefont
  {de~Lange}}, \bibinfo {author} {\bibfnamefont {Z.~H.}\ \bibnamefont {Wang}},
  \bibinfo {author} {\bibfnamefont {D.}~\bibnamefont {Rist{\`e}}}, \bibinfo
  {author} {\bibfnamefont {V.~V.}\ \bibnamefont {Dobrovitski}}, \ and\ \bibinfo
  {author} {\bibfnamefont {R.}~\bibnamefont {Hanson}},\ }\bibfield  {title}
  {\enquote {\bibinfo {title} {Universal \text{D}ynamical \text{D}ecoupling of
  a \text{S}ingle \text{S}olid-\text{S}tate \text{S}pin from a \text{S}pin
  \text{B}ath},}\ }\href {\doibase 10.1126/science.1192739} {\bibfield
  {journal} {\bibinfo  {journal} {Science}\ }\textbf {\bibinfo {volume}
  {330}},\ \bibinfo {pages} {60} (\bibinfo {year} {2010})}\BibitemShut
  {NoStop}%
\bibitem [{\citenamefont {Naydenov}\ \emph {et~al.}(2011)\citenamefont
  {Naydenov}, \citenamefont {Dolde}, \citenamefont {Hall}, \citenamefont
  {Shin}, \citenamefont {Fedder}, \citenamefont {Hollenberg}, \citenamefont
  {Jelezko},\ and\ \citenamefont {Wrachtrup}}]{Naydenov2011}%
  \BibitemOpen
  \bibfield  {author} {\bibinfo {author} {\bibfnamefont {B.}~\bibnamefont
  {Naydenov}}, \bibinfo {author} {\bibfnamefont {F.}~\bibnamefont {Dolde}},
  \bibinfo {author} {\bibfnamefont {L.~T.}\ \bibnamefont {Hall}}, \bibinfo
  {author} {\bibfnamefont {C.}~\bibnamefont {Shin}}, \bibinfo {author}
  {\bibfnamefont {H.}~\bibnamefont {Fedder}}, \bibinfo {author} {\bibfnamefont
  {L.~C.~L.}\ \bibnamefont {Hollenberg}}, \bibinfo {author} {\bibfnamefont
  {F.}~\bibnamefont {Jelezko}}, \ and\ \bibinfo {author} {\bibfnamefont
  {J.}~\bibnamefont {Wrachtrup}},\ }\bibfield  {title} {\enquote {\bibinfo
  {title} {Dynamical decoupling of a single-electron spin at room
  temperature},}\ }\href {\doibase 10.1103/PhysRevB.83.081201} {\bibfield
  {journal} {\bibinfo  {journal} {Phys. Rev. B}\ }\textbf {\bibinfo {volume}
  {83}},\ \bibinfo {pages} {081201} (\bibinfo {year} {2011})}\BibitemShut
  {NoStop}%
\bibitem [{\citenamefont {Taminiau}\ \emph {et~al.}(2012)\citenamefont
  {Taminiau}, \citenamefont {Wagenaar}, \citenamefont {van~der Sar},
  \citenamefont {Jelezko}, \citenamefont {Dobrovitski},\ and\ \citenamefont
  {Hanson}}]{Taminiau2012}%
  \BibitemOpen
  \bibfield  {author} {\bibinfo {author} {\bibfnamefont {T.~H.}\ \bibnamefont
  {Taminiau}}, \bibinfo {author} {\bibfnamefont {J.~J.~T.}\ \bibnamefont
  {Wagenaar}}, \bibinfo {author} {\bibfnamefont {T.}~\bibnamefont {van~der
  Sar}}, \bibinfo {author} {\bibfnamefont {F.}~\bibnamefont {Jelezko}},
  \bibinfo {author} {\bibfnamefont {V.~V.}\ \bibnamefont {Dobrovitski}}, \ and\
  \bibinfo {author} {\bibfnamefont {R.}~\bibnamefont {Hanson}},\ }\bibfield
  {title} {\enquote {\bibinfo {title} {Detection and \text{C}ontrol of
  \text{I}ndividual \text{N}uclear \text{S}pins \text{U}sing a \text{W}eakly
  \text{C}oupled \text{E}lectron \text{S}pin},}\ }\href {\doibase
  10.1103/PhysRevLett.109.137602} {\bibfield  {journal} {\bibinfo  {journal}
  {Phys. Rev. Lett.}\ }\textbf {\bibinfo {volume} {109}},\ \bibinfo {pages}
  {137602} (\bibinfo {year} {2012})}\BibitemShut {NoStop}%
\bibitem [{\citenamefont {Zhao}\ \emph {et~al.}(2011)\citenamefont {Zhao},
  \citenamefont {Hu}, \citenamefont {Ho}, \citenamefont {Wan},\ and\
  \citenamefont {Liu}}]{Zhao2011}%
  \BibitemOpen
  \bibfield  {author} {\bibinfo {author} {\bibfnamefont {N.}~\bibnamefont
  {Zhao}}, \bibinfo {author} {\bibfnamefont {J.~L.}\ \bibnamefont {Hu}},
  \bibinfo {author} {\bibfnamefont {S.-W.}\ \bibnamefont {Ho}}, \bibinfo
  {author} {\bibfnamefont {J.~T.}\ \bibnamefont {Wan}}, \ and\ \bibinfo
  {author} {\bibfnamefont {R.~B.}\ \bibnamefont {Liu}},\ }\bibfield  {title}
  {\enquote {\bibinfo {title} {Atomic-scale magnetometry of distant nuclear
  spin clusters via nitrogen-vacancy spin in diamond},}\ }\href {\doibase
  10.1038/nnano.2011.22} {\bibfield  {journal} {\bibinfo  {journal} {Nat.
  Nanotechnol.}\ }\textbf {\bibinfo {volume} {6}},\ \bibinfo {pages} {242}
  (\bibinfo {year} {2011})}\BibitemShut {NoStop}%
\bibitem [{\citenamefont {Kolkowitz}\ \emph {et~al.}(2012)\citenamefont
  {Kolkowitz}, \citenamefont {Unterreithmeier}, \citenamefont {Bennett},\ and\
  \citenamefont {Lukin}}]{Kolkowitz2012}%
  \BibitemOpen
  \bibfield  {author} {\bibinfo {author} {\bibfnamefont {S.}~\bibnamefont
  {Kolkowitz}}, \bibinfo {author} {\bibfnamefont {Q.~P.}\ \bibnamefont
  {Unterreithmeier}}, \bibinfo {author} {\bibfnamefont {S.~D.}\ \bibnamefont
  {Bennett}}, \ and\ \bibinfo {author} {\bibfnamefont {M.~D.}\ \bibnamefont
  {Lukin}},\ }\bibfield  {title} {\enquote {\bibinfo {title} {Sensing
  \text{D}istant \text{N}uclear \text{S}pins with a \text{S}ingle
  \text{E}lectron \text{S}pin},}\ }\href {\doibase
  10.1103/PhysRevLett.109.137601} {\bibfield  {journal} {\bibinfo  {journal}
  {Phys. Rev. Lett.}\ }\textbf {\bibinfo {volume} {109}},\ \bibinfo {pages}
  {137601} (\bibinfo {year} {2012})}\BibitemShut {NoStop}%
\bibitem [{\citenamefont {Liu}\ \emph {et~al.}(2017)\citenamefont {Liu},
  \citenamefont {Xing}, \citenamefont {Ma}, \citenamefont {Wang}, \citenamefont
  {Li}, \citenamefont {Po}, \citenamefont {Zhang}, \citenamefont {Fan},
  \citenamefont {Liu},\ and\ \citenamefont {Pan}}]{Liu2017}%
  \BibitemOpen
  \bibfield  {author} {\bibinfo {author} {\bibfnamefont {G.~Q.}\ \bibnamefont
  {Liu}}, \bibinfo {author} {\bibfnamefont {J.}~\bibnamefont {Xing}}, \bibinfo
  {author} {\bibfnamefont {W.~L.}\ \bibnamefont {Ma}}, \bibinfo {author}
  {\bibfnamefont {P.}~\bibnamefont {Wang}}, \bibinfo {author} {\bibfnamefont
  {C.~H.}\ \bibnamefont {Li}}, \bibinfo {author} {\bibfnamefont {H.~C.}\
  \bibnamefont {Po}}, \bibinfo {author} {\bibfnamefont {Y.~R.}\ \bibnamefont
  {Zhang}}, \bibinfo {author} {\bibfnamefont {H.}~\bibnamefont {Fan}}, \bibinfo
  {author} {\bibfnamefont {R.~B.}\ \bibnamefont {Liu}}, \ and\ \bibinfo
  {author} {\bibfnamefont {X.~Y.}\ \bibnamefont {Pan}},\ }\bibfield  {title}
  {\enquote {\bibinfo {title} {Single-\text{S}hot \text{R}eadout of a
  \text{N}uclear \text{S}pin \text{W}eakly \text{C}oupled to a
  \text{N}itrogen-\text{V}acancy \text{C}enter at \text{R}oom
  \text{T}emperature},}\ }\href {\doibase 10.1103/PhysRevLett.118.150504}
  {\bibfield  {journal} {\bibinfo  {journal} {Phys. Rev. Lett}\ }\textbf
  {\bibinfo {volume} {118}},\ \bibinfo {pages} {150504} (\bibinfo {year}
  {2017})}\BibitemShut {NoStop}%
\bibitem [{\citenamefont {Johanning}\ \emph {et~al.}(2009)\citenamefont
  {Johanning}, \citenamefont {Braun}, \citenamefont {Timoney}, \citenamefont
  {Elman}, \citenamefont {Neuhauser},\ and\ \citenamefont
  {Wunderlich}}]{Johanning2009}%
  \BibitemOpen
  \bibfield  {author} {\bibinfo {author} {\bibfnamefont {M.}~\bibnamefont
  {Johanning}}, \bibinfo {author} {\bibfnamefont {A.}~\bibnamefont {Braun}},
  \bibinfo {author} {\bibfnamefont {N.}~\bibnamefont {Timoney}}, \bibinfo
  {author} {\bibfnamefont {V.}~\bibnamefont {Elman}}, \bibinfo {author}
  {\bibfnamefont {W.}~\bibnamefont {Neuhauser}}, \ and\ \bibinfo {author}
  {\bibfnamefont {Chr.}\ \bibnamefont {Wunderlich}},\ }\bibfield  {title}
  {\enquote {\bibinfo {title} {Individual \text{A}ddressing of \text{T}rapped
  \text{I}ons and \text{C}oupling of \text{M}otional and \text{S}pin
  \text{S}tates \text{U}sing rf \text{R}adiation},}\ }\href {\doibase
  10.1103/PhysRevLett.102.073004} {\bibfield  {journal} {\bibinfo  {journal}
  {Phys. Rev. Lett.}\ }\textbf {\bibinfo {volume} {102}},\ \bibinfo {pages}
  {073004} (\bibinfo {year} {2009})}\BibitemShut {NoStop}%
\bibitem [{\citenamefont {Khromova}\ \emph {et~al.}(2012)\citenamefont
  {Khromova}, \citenamefont {Piltz}, \citenamefont {Scharfenberger},
  \citenamefont {Gloger}, \citenamefont {Johanning}, \citenamefont {Var\'on},\
  and\ \citenamefont {Wunderlich}}]{Khromova2012}%
  \BibitemOpen
  \bibfield  {author} {\bibinfo {author} {\bibfnamefont {A.}~\bibnamefont
  {Khromova}}, \bibinfo {author} {\bibfnamefont {Ch.}\ \bibnamefont {Piltz}},
  \bibinfo {author} {\bibfnamefont {B.}~\bibnamefont {Scharfenberger}},
  \bibinfo {author} {\bibfnamefont {T.~F.}\ \bibnamefont {Gloger}}, \bibinfo
  {author} {\bibfnamefont {M.}~\bibnamefont {Johanning}}, \bibinfo {author}
  {\bibfnamefont {A.~F.}\ \bibnamefont {Var\'on}}, \ and\ \bibinfo {author}
  {\bibfnamefont {Ch.}\ \bibnamefont {Wunderlich}},\ }\bibfield  {title}
  {\enquote {\bibinfo {title} {Designer \text{S}pin \text{P}seudomolecule
  \text{I}mplemented with \text{T}rapped \text{I}ons in a \text{M}agnetic
  \text{G}radient},}\ }\href {\doibase 10.1103/PhysRevLett.108.220502}
  {\bibfield  {journal} {\bibinfo  {journal} {Phys. Rev. Lett.}\ }\textbf
  {\bibinfo {volume} {108}},\ \bibinfo {pages} {220502} (\bibinfo {year}
  {2012})}\BibitemShut {NoStop}%
\bibitem [{\citenamefont {Sriarunothai}\ \emph {et~al.}(2018)\citenamefont
  {Sriarunothai}, \citenamefont {Giri}, \citenamefont {W\"olk},\ and\
  \citenamefont {Wunderlich}}]{Sriarunothai2017}%
  \BibitemOpen
  \bibfield  {author} {\bibinfo {author} {\bibfnamefont {T.}~\bibnamefont
  {Sriarunothai}}, \bibinfo {author} {\bibfnamefont {G.~S.}\ \bibnamefont
  {Giri}}, \bibinfo {author} {\bibfnamefont {S.}~\bibnamefont {W\"olk}}, \ and\
  \bibinfo {author} {\bibfnamefont {Ch.}\ \bibnamefont {Wunderlich}},\
  }\bibfield  {title} {\enquote {\bibinfo {title} {Radio frequency sideband
  cooling and sympathetic cooling of trapped ions in a static magnetic field
  gradient},}\ }\href {\doibase 10.1080/09500340.2017.1401137} {\bibfield
  {journal} {\bibinfo  {journal} {J. Mod. Opt.}\ }\textbf {\bibinfo {volume}
  {65}},\ \bibinfo {pages} {560} (\bibinfo {year} {2018})}\BibitemShut
  {NoStop}%
\bibitem [{\citenamefont {Li}\ \emph {et~al.}(2011)\citenamefont {Li},
  \citenamefont {Ruths}, \citenamefont {Yu}, \citenamefont {Arthanari},\ and\
  \citenamefont {Wagner}}]{Li2011}%
  \BibitemOpen
  \bibfield  {author} {\bibinfo {author} {\bibfnamefont {J.-S.}\ \bibnamefont
  {Li}}, \bibinfo {author} {\bibfnamefont {J.}~\bibnamefont {Ruths}}, \bibinfo
  {author} {\bibfnamefont {T.-Y.}\ \bibnamefont {Yu}}, \bibinfo {author}
  {\bibfnamefont {H.}~\bibnamefont {Arthanari}}, \ and\ \bibinfo {author}
  {\bibfnamefont {G.}~\bibnamefont {Wagner}},\ }\bibfield  {title} {\enquote
  {\bibinfo {title} {Optimal pulse design in quantum control: A unified
  computational method},}\ }\href {\doibase 10.1073/pnas.1009797108} {\bibfield
   {journal} {\bibinfo  {journal} {Proc. Natl. Acad. Sci. U.S.A.}\ }\textbf
  {\bibinfo {volume} {108}},\ \bibinfo {pages} {1879} (\bibinfo {year}
  {2011})}\BibitemShut {NoStop}%
\bibitem [{\citenamefont {Zhao}\ \emph {et~al.}(2014)\citenamefont {Zhao},
  \citenamefont {Wrachtrup},\ and\ \citenamefont {Liu}}]{Zhao2014}%
  \BibitemOpen
  \bibfield  {author} {\bibinfo {author} {\bibfnamefont {N.}~\bibnamefont
  {Zhao}}, \bibinfo {author} {\bibfnamefont {J.}~\bibnamefont {Wrachtrup}}, \
  and\ \bibinfo {author} {\bibfnamefont {R.~B.}\ \bibnamefont {Liu}},\
  }\bibfield  {title} {\enquote {\bibinfo {title} {Dynamical decoupling design
  for identifying weakly coupled nuclear spins in a bath},}\ }\href {\doibase
  10.1103/PhysRevA.90.032319} {\bibfield  {journal} {\bibinfo  {journal} {Phys.
  Rev. A}\ }\textbf {\bibinfo {volume} {90}},\ \bibinfo {pages} {032319}
  (\bibinfo {year} {2014})}\BibitemShut {NoStop}%
\bibitem [{Sup()}]{Supp}%
  \BibitemOpen
  \href@noop {} {}\bibinfo {note} {See Supplemental Material for more details,
  including
  Refs.~\cite{Kolkowitz2012,Taminiau2012,Grinolds2014,Zopes2018,Dobrovitski2008,Dobrovitski2009,de-Lange2010,Maze2012,Wang2013,VanKampen2007,Gillespie1991,Gillespie1995,Gillespie1996,Naydenov2011,Cahill1969a}.}\BibitemShut
  {Stop}%
\bibitem [{\citenamefont {Albertini}\ and\ \citenamefont
  {D'Alessandro}(2002)}]{Albertini2002}%
  \BibitemOpen
  \bibfield  {author} {\bibinfo {author} {\bibfnamefont {F.}~\bibnamefont
  {Albertini}}\ and\ \bibinfo {author} {\bibfnamefont {D.}~\bibnamefont
  {D'Alessandro}},\ }\bibfield  {title} {\enquote {\bibinfo {title} {The
  \text{L}ie algebra structure and controllability of spin systems},}\ }\href
  {\doibase 10.1016/S0024-3795(02)00290-2} {\bibfield  {journal} {\bibinfo
  {journal} {Linear Algebra Appl.}\ }\textbf {\bibinfo {volume} {350}},\
  \bibinfo {pages} {213} (\bibinfo {year} {2002})}\BibitemShut {NoStop}%
\bibitem [{\citenamefont {Cai}\ \emph {et~al.}(2013)\citenamefont {Cai},
  \citenamefont {Retzker}, \citenamefont {Jelezko},\ and\ \citenamefont
  {Plenio}}]{Cai2013}%
  \BibitemOpen
  \bibfield  {author} {\bibinfo {author} {\bibfnamefont {J.-M.}\ \bibnamefont
  {Cai}}, \bibinfo {author} {\bibfnamefont {A.}~\bibnamefont {Retzker}},
  \bibinfo {author} {\bibfnamefont {F.}~\bibnamefont {Jelezko}}, \ and\
  \bibinfo {author} {\bibfnamefont {M.~B.}\ \bibnamefont {Plenio}},\ }\bibfield
   {title} {\enquote {\bibinfo {title} {A large-scale quantum simulator on a
  diamond surface at room temperature},}\ }\href {\doibase 10.1038/nphys2519}
  {\bibfield  {journal} {\bibinfo  {journal} {Nat. Phys.}\ }\textbf {\bibinfo
  {volume} {9}},\ \bibinfo {pages} {168} (\bibinfo {year} {2013})}\BibitemShut
  {NoStop}%
\bibitem [{\citenamefont {Chen}\ \emph {et~al.}(2013)\citenamefont {Chen},
  \citenamefont {Jiang},\ and\ \citenamefont {Liu}}]{Chen2013}%
  \BibitemOpen
  \bibfield  {author} {\bibinfo {author} {\bibfnamefont {S.-W.}\ \bibnamefont
  {Chen}}, \bibinfo {author} {\bibfnamefont {Z.-F.}\ \bibnamefont {Jiang}}, \
  and\ \bibinfo {author} {\bibfnamefont {R.-B.}\ \bibnamefont {Liu}},\
  }\bibfield  {title} {\enquote {\bibinfo {title} {Quantum criticality at high
  temperature revealed by spin echo},}\ }\href {\doibase
  10.1088/1367-2630/15/4/043032} {\bibfield  {journal} {\bibinfo  {journal}
  {New J. Phys.}\ }\textbf {\bibinfo {volume} {15}},\ \bibinfo {pages} {043032}
  (\bibinfo {year} {2013})}\BibitemShut {NoStop}%
\bibitem [{\citenamefont {Xu}\ and\ \citenamefont {del Campo}()}]{Xu2018}%
  \BibitemOpen
  \bibfield  {author} {\bibinfo {author} {\bibfnamefont {Z.}~\bibnamefont
  {Xu}}\ and\ \bibinfo {author} {\bibfnamefont {A.}~\bibnamefont {del Campo}},\
  }\bibfield  {title} {\enquote {\bibinfo {title} {Probing the full
  distribution of many-body observables by single-qubit interferometry},}\
  }\href {https://arxiv.org/abs/1812.06983} {\bibinfo  {journal} {1812.06983}\
  }\BibitemShut {NoStop}%
\bibitem [{\citenamefont {Cahill}\ and\ \citenamefont
  {Glauber}(1969)}]{Cahill1969a}%
  \BibitemOpen
\bibfield  {journal} {  }\bibfield  {author} {\bibinfo {author} {\bibfnamefont
  {K.~E.}\ \bibnamefont {Cahill}}\ and\ \bibinfo {author} {\bibfnamefont
  {R.~J.}\ \bibnamefont {Glauber}},\ }\bibfield  {title} {\enquote {\bibinfo
  {title} {Ordered \text{E}xpansions in \text{B}oson \text{A}mplitude
  \text{O}perators},}\ }\href {\doibase 10.1103/PhysRev.177.1857} {\bibfield
  {journal} {\bibinfo  {journal} {Phys. Rev.}\ }\textbf {\bibinfo {volume}
  {177}},\ \bibinfo {pages} {1857} (\bibinfo {year} {1969})}\BibitemShut
  {NoStop}%
\bibitem [{\citenamefont {Wigner}(1932)}]{Wigner1932}%
  \BibitemOpen
  \bibfield  {author} {\bibinfo {author} {\bibfnamefont {E.}~\bibnamefont
  {Wigner}},\ }\bibfield  {title} {\enquote {\bibinfo {title} {On the
  \text{Q}uantum \text{C}orrection \text{F}or \text{T}hermodynamic
  \text{E}quilibrium},}\ }\href {\doibase 10.1103/PhysRev.40.749} {\bibfield
  {journal} {\bibinfo  {journal} {Phys. Rev.}\ }\textbf {\bibinfo {volume}
  {40}},\ \bibinfo {pages} {749} (\bibinfo {year} {1932})}\BibitemShut
  {NoStop}%
\bibitem [{\citenamefont {Doherty}\ \emph {et~al.}(2013)\citenamefont
  {Doherty}, \citenamefont {Manson}, \citenamefont {Delaney}, \citenamefont
  {Jelezko}, \citenamefont {Wrachtrup},\ and\ \citenamefont
  {Hollenberg}}]{Doherty2013}%
  \BibitemOpen
  \bibfield  {author} {\bibinfo {author} {\bibfnamefont {M.~W.}\ \bibnamefont
  {Doherty}}, \bibinfo {author} {\bibfnamefont {N.~B.}\ \bibnamefont {Manson}},
  \bibinfo {author} {\bibfnamefont {P.}~\bibnamefont {Delaney}}, \bibinfo
  {author} {\bibfnamefont {F.}~\bibnamefont {Jelezko}}, \bibinfo {author}
  {\bibfnamefont {J.}~\bibnamefont {Wrachtrup}}, \ and\ \bibinfo {author}
  {\bibfnamefont {C.~L.~C.}\ \bibnamefont {Hollenberg}},\ }\bibfield  {title}
  {\enquote {\bibinfo {title} {The nitrogen-vacancy colour centre in
  diamond},}\ }\href {\doibase 10.1016/j.physrep.2013.02.001} {\bibfield
  {journal} {\bibinfo  {journal} {Phys. Rep.}\ }\textbf {\bibinfo {volume}
  {528}},\ \bibinfo {pages} {1} (\bibinfo {year} {2013})}\BibitemShut {NoStop}%
\bibitem [{\citenamefont {Robledo}\ \emph {et~al.}(2010)\citenamefont
  {Robledo}, \citenamefont {Bernien}, \citenamefont {van Weperen},\ and\
  \citenamefont {Hanson}}]{Robledo2010}%
  \BibitemOpen
  \bibfield  {author} {\bibinfo {author} {\bibfnamefont {L.}~\bibnamefont
  {Robledo}}, \bibinfo {author} {\bibfnamefont {H.}~\bibnamefont {Bernien}},
  \bibinfo {author} {\bibfnamefont {I.}~\bibnamefont {van Weperen}}, \ and\
  \bibinfo {author} {\bibfnamefont {R.}~\bibnamefont {Hanson}},\ }\bibfield
  {title} {\enquote {\bibinfo {title} {Control and \text{C}oherence of the
  \text{O}ptical \text{T}ransition of \text{S}ingle \text{N}itrogen
  \text{V}acancy \text{C}enters in \text{D}iamond},}\ }\href {\doibase
  10.1103/PhysRevLett.105.177403} {\bibfield  {journal} {\bibinfo  {journal}
  {Phys. Rev. Lett.}\ }\textbf {\bibinfo {volume} {105}},\ \bibinfo {pages}
  {177403} (\bibinfo {year} {2010})}\BibitemShut {NoStop}%
\bibitem [{\citenamefont {de~Lange}\ \emph {et~al.}(2011)\citenamefont
  {de~Lange}, \citenamefont {Rist{\`e}}, \citenamefont {Dobrovitski},\ and\
  \citenamefont {Hanson}}]{de-Lange2011}%
  \BibitemOpen
  \bibfield  {author} {\bibinfo {author} {\bibfnamefont {G.}~\bibnamefont
  {de~Lange}}, \bibinfo {author} {\bibfnamefont {D.}~\bibnamefont {Rist{\`e}}},
  \bibinfo {author} {\bibfnamefont {V.~V.}\ \bibnamefont {Dobrovitski}}, \ and\
  \bibinfo {author} {\bibfnamefont {R.}~\bibnamefont {Hanson}},\ }\bibfield
  {title} {\enquote {\bibinfo {title} {Single-\text{S}pin \text{M}agnetometry
  with \text{M}ultipulse \text{S}ensing \text{S}equences},}\ }\href {\doibase
  10.1103/PhysRevLett.106.080802} {\bibfield  {journal} {\bibinfo  {journal}
  {Phys. Rev. Lett.}\ }\textbf {\bibinfo {volume} {106}},\ \bibinfo {pages}
  {080802} (\bibinfo {year} {2011})}\BibitemShut {NoStop}%
\bibitem [{\citenamefont {Ryan}\ \emph {et~al.}(2010)\citenamefont {Ryan},
  \citenamefont {Hodges},\ and\ \citenamefont {Cory}}]{Ryan2010}%
  \BibitemOpen
  \bibfield  {author} {\bibinfo {author} {\bibfnamefont {C.~A.}\ \bibnamefont
  {Ryan}}, \bibinfo {author} {\bibfnamefont {J.~S.}\ \bibnamefont {Hodges}}, \
  and\ \bibinfo {author} {\bibfnamefont {D.~G.}\ \bibnamefont {Cory}},\
  }\bibfield  {title} {\enquote {\bibinfo {title} {Robust \text{R}ecoupling
  \text{T}echniques to \text{E}xtend \text{Q}uantum \text{C}oherence in
  \text{D}iamond},}\ }\href {\doibase 10.1103/PhysRevLett.105.200402}
  {\bibfield  {journal} {\bibinfo  {journal} {Phys. Rev. Lett.}\ }\textbf
  {\bibinfo {volume} {105}},\ \bibinfo {pages} {200402} (\bibinfo {year}
  {2010})}\BibitemShut {NoStop}%
\bibitem [{\citenamefont {Bar-Gill}\ \emph {et~al.}(2013)\citenamefont
  {Bar-Gill}, \citenamefont {Pham}, \citenamefont {Jarmola}, \citenamefont
  {Budker},\ and\ \citenamefont {Walsworth}}]{Bar-Gill2013}%
  \BibitemOpen
  \bibfield  {author} {\bibinfo {author} {\bibfnamefont {N.}~\bibnamefont
  {Bar-Gill}}, \bibinfo {author} {\bibfnamefont {L.~M.}\ \bibnamefont {Pham}},
  \bibinfo {author} {\bibfnamefont {A.}~\bibnamefont {Jarmola}}, \bibinfo
  {author} {\bibfnamefont {D.}~\bibnamefont {Budker}}, \ and\ \bibinfo {author}
  {\bibfnamefont {R.~L.}\ \bibnamefont {Walsworth}},\ }\bibfield  {title}
  {\enquote {\bibinfo {title} {Solid-state electronic spin coherence time
  approaching one second},}\ }\href {\doibase 10.1038/ncomms2771} {\bibfield
  {journal} {\bibinfo  {journal} {Nat. Commun.}\ }\textbf {\bibinfo {volume}
  {4}},\ \bibinfo {pages} {1743} (\bibinfo {year} {2013})}\BibitemShut
  {NoStop}%
\bibitem [{\citenamefont {Fisk}\ \emph {et~al.}(1997)\citenamefont {Fisk},
  \citenamefont {Sellars}, \citenamefont {Lawn},\ and\ \citenamefont
  {Coles}}]{Fisk1997}%
  \BibitemOpen
  \bibfield  {author} {\bibinfo {author} {\bibfnamefont {P.~T.~H.}\
  \bibnamefont {Fisk}}, \bibinfo {author} {\bibfnamefont {M.~J.}\ \bibnamefont
  {Sellars}}, \bibinfo {author} {\bibfnamefont {M.~A.}\ \bibnamefont {Lawn}}, \
  and\ \bibinfo {author} {\bibfnamefont {G.}~\bibnamefont {Coles}},\ }\bibfield
   {title} {\enquote {\bibinfo {title} {Accurate measurement of the 12.6
  \text{GH}z "clock" transition in trapped $^{171}$\text{Y}b$^+$ ions},}\
  }\href {\doibase 10.1109/58.585119} {\bibfield  {journal} {\bibinfo
  {journal} {IEEE Trans. Ultrason. Ferroelectr. Freq. Control}\ }\textbf
  {\bibinfo {volume} {44}},\ \bibinfo {pages} {344} (\bibinfo {year}
  {1997})}\BibitemShut {NoStop}%
\bibitem [{\citenamefont {Campbell}\ \emph {et~al.}(2010)\citenamefont
  {Campbell}, \citenamefont {Mizrahi}, \citenamefont {Quraishi}, \citenamefont
  {Senko}, \citenamefont {Hayes}, \citenamefont {Hucul}, \citenamefont
  {Matsukevich}, \citenamefont {Maunz},\ and\ \citenamefont
  {Monroe}}]{Campbell2010}%
  \BibitemOpen
  \bibfield  {author} {\bibinfo {author} {\bibfnamefont {W.~C.}\ \bibnamefont
  {Campbell}}, \bibinfo {author} {\bibfnamefont {J.}~\bibnamefont {Mizrahi}},
  \bibinfo {author} {\bibfnamefont {Q.}~\bibnamefont {Quraishi}}, \bibinfo
  {author} {\bibfnamefont {C.}~\bibnamefont {Senko}}, \bibinfo {author}
  {\bibfnamefont {D.}~\bibnamefont {Hayes}}, \bibinfo {author} {\bibfnamefont
  {D.}~\bibnamefont {Hucul}}, \bibinfo {author} {\bibfnamefont {D.~N.}\
  \bibnamefont {Matsukevich}}, \bibinfo {author} {\bibfnamefont
  {P.}~\bibnamefont {Maunz}}, \ and\ \bibinfo {author} {\bibfnamefont
  {C.}~\bibnamefont {Monroe}},\ }\bibfield  {title} {\enquote {\bibinfo {title}
  {Ultrafast \text{G}ates for \text{S}ingle \text{A}tomic \text{Q}ubits},}\
  }\href {\doibase 10.1103/PhysRevLett.105.090502} {\bibfield  {journal}
  {\bibinfo  {journal} {Phys. Rev. Lett.}\ }\textbf {\bibinfo {volume} {105}},\
  \bibinfo {pages} {090502} (\bibinfo {year} {2010})}\BibitemShut {NoStop}%
\bibitem [{\citenamefont {Grinolds}\ \emph {et~al.}(2014)\citenamefont
  {Grinolds}, \citenamefont {Warner}, \citenamefont {De~Greve}, \citenamefont
  {Dovzhenko}, \citenamefont {Thiel}, \citenamefont {Walsworth}, \citenamefont
  {Hong}, \citenamefont {Maletinsky},\ and\ \citenamefont
  {Yacoby}}]{Grinolds2014}%
  \BibitemOpen
  \bibfield  {author} {\bibinfo {author} {\bibfnamefont {M.~S.}\ \bibnamefont
  {Grinolds}}, \bibinfo {author} {\bibfnamefont {M.}~\bibnamefont {Warner}},
  \bibinfo {author} {\bibfnamefont {K.}~\bibnamefont {De~Greve}}, \bibinfo
  {author} {\bibfnamefont {Y.}~\bibnamefont {Dovzhenko}}, \bibinfo {author}
  {\bibfnamefont {L.}~\bibnamefont {Thiel}}, \bibinfo {author} {\bibfnamefont
  {R.~L.}\ \bibnamefont {Walsworth}}, \bibinfo {author} {\bibfnamefont
  {S.}~\bibnamefont {Hong}}, \bibinfo {author} {\bibfnamefont {P.}~\bibnamefont
  {Maletinsky}}, \ and\ \bibinfo {author} {\bibfnamefont {A.}~\bibnamefont
  {Yacoby}},\ }\bibfield  {title} {\enquote {\bibinfo {title} {Subnanometre
  resolution in three-dimensional magnetic resonance imaging of individual dark
  spins},}\ }\href {\doibase 10.1038/NNANO.2014.30} {\bibfield  {journal}
  {\bibinfo  {journal} {Nat. Nanotechnol.}\ }\textbf {\bibinfo {volume} {9}},\
  \bibinfo {pages} {279} (\bibinfo {year} {2014})}\BibitemShut {NoStop}%
\bibitem [{\citenamefont {Zopes}\ \emph {et~al.}(2018)\citenamefont {Zopes},
  \citenamefont {Cujia}, \citenamefont {Sasaki}, \citenamefont {Boss},
  \citenamefont {Itoh},\ and\ \citenamefont {Degen}}]{Zopes2018}%
  \BibitemOpen
  \bibfield  {author} {\bibinfo {author} {\bibfnamefont {J.}~\bibnamefont
  {Zopes}}, \bibinfo {author} {\bibfnamefont {K.~S.}\ \bibnamefont {Cujia}},
  \bibinfo {author} {\bibfnamefont {K.}~\bibnamefont {Sasaki}}, \bibinfo
  {author} {\bibfnamefont {J.~M.}\ \bibnamefont {Boss}}, \bibinfo {author}
  {\bibfnamefont {K.~M.}\ \bibnamefont {Itoh}}, \ and\ \bibinfo {author}
  {\bibfnamefont {C.~L.}\ \bibnamefont {Degen}},\ }\bibfield  {title} {\enquote
  {\bibinfo {title} {Three-dimensional localization spectroscopy of individual
  nuclear spins with sub-angstrom resolution},}\ }\href {\doibase
  10.1038/s41467-018-07121-0} {\bibfield  {journal} {\bibinfo  {journal} {Nat.
  Commun.}\ }\textbf {\bibinfo {volume} {9}},\ \bibinfo {pages} {4678}
  (\bibinfo {year} {2018})}\BibitemShut {NoStop}%
\bibitem [{\citenamefont {Dobrovitski}\ \emph {et~al.}(2008)\citenamefont
  {Dobrovitski}, \citenamefont {Feiguin}, \citenamefont {Awschalom},\ and\
  \citenamefont {Hanson}}]{Dobrovitski2008}%
  \BibitemOpen
  \bibfield  {author} {\bibinfo {author} {\bibfnamefont {V.~V.}\ \bibnamefont
  {Dobrovitski}}, \bibinfo {author} {\bibfnamefont {A.~E.}\ \bibnamefont
  {Feiguin}}, \bibinfo {author} {\bibfnamefont {D.~D.}\ \bibnamefont
  {Awschalom}}, \ and\ \bibinfo {author} {\bibfnamefont {R.}~\bibnamefont
  {Hanson}},\ }\bibfield  {title} {\enquote {\bibinfo {title} {Decoherence
  dynamics of a single spin versus spin ensemble},}\ }\href {\doibase
  10.1103/PhysRevB.77.245212} {\bibfield  {journal} {\bibinfo  {journal} {Phys.
  Rev. B}\ }\textbf {\bibinfo {volume} {77}},\ \bibinfo {pages} {245212}
  (\bibinfo {year} {2008})}\BibitemShut {NoStop}%
\bibitem [{\citenamefont {Dobrovitski}\ \emph {et~al.}(2009)\citenamefont
  {Dobrovitski}, \citenamefont {Feiguin}, \citenamefont {Hanson},\ and\
  \citenamefont {Awschalom}}]{Dobrovitski2009}%
  \BibitemOpen
  \bibfield  {author} {\bibinfo {author} {\bibfnamefont {V.~V.}\ \bibnamefont
  {Dobrovitski}}, \bibinfo {author} {\bibfnamefont {A.~E.}\ \bibnamefont
  {Feiguin}}, \bibinfo {author} {\bibfnamefont {R.}~\bibnamefont {Hanson}}, \
  and\ \bibinfo {author} {\bibfnamefont {D.~D.}\ \bibnamefont {Awschalom}},\
  }\bibfield  {title} {\enquote {\bibinfo {title} {Decay of \text{R}abi
  \text{O}scillations by \text{D}ipolar-\text{C}oupled \text{D}ynamical
  \text{S}pin \text{E}nvironments},}\ }\href {\doibase
  10.1103/PhysRevLett.102.237601} {\bibfield  {journal} {\bibinfo  {journal}
  {Phys. Rev. Lett.}\ }\textbf {\bibinfo {volume} {102}},\ \bibinfo {pages}
  {237601} (\bibinfo {year} {2009})}\BibitemShut {NoStop}%
\bibitem [{\citenamefont {Maze}\ \emph {et~al.}(2012)\citenamefont {Maze},
  \citenamefont {Dr\'eau}, \citenamefont {Waselowski}, \citenamefont {Duarte},
  \citenamefont {Roch},\ and\ \citenamefont {Jacques}}]{Maze2012}%
  \BibitemOpen
  \bibfield  {author} {\bibinfo {author} {\bibfnamefont {J.~R.}\ \bibnamefont
  {Maze}}, \bibinfo {author} {\bibfnamefont {A.}~\bibnamefont {Dr\'eau}},
  \bibinfo {author} {\bibfnamefont {V.}~\bibnamefont {Waselowski}}, \bibinfo
  {author} {\bibfnamefont {H.}~\bibnamefont {Duarte}}, \bibinfo {author}
  {\bibfnamefont {J.-F.}\ \bibnamefont {Roch}}, \ and\ \bibinfo {author}
  {\bibfnamefont {V}~\bibnamefont {Jacques}},\ }\bibfield  {title} {\enquote
  {\bibinfo {title} {Free induction decay of single spins in diamond},}\ }\href
  {\doibase 10.1088/1367-2630/14/10/103041} {\bibfield  {journal} {\bibinfo
  {journal} {New J. Phys.}\ }\textbf {\bibinfo {volume} {14}},\ \bibinfo
  {pages} {103041} (\bibinfo {year} {2012})}\BibitemShut {NoStop}%
\bibitem [{\citenamefont {Wang}\ and\ \citenamefont
  {Takahashi}(2013)}]{Wang2013}%
  \BibitemOpen
  \bibfield  {author} {\bibinfo {author} {\bibfnamefont {Z.-H.}\ \bibnamefont
  {Wang}}\ and\ \bibinfo {author} {\bibfnamefont {S.}~\bibnamefont
  {Takahashi}},\ }\bibfield  {title} {\enquote {\bibinfo {title} {Spin
  decoherence and electron spin bath noise of a nitrogen-vacancy center in
  diamond},}\ }\href {\doibase 10.1103/PhysRevB.87.115122} {\bibfield
  {journal} {\bibinfo  {journal} {Phys. Rev. B}\ }\textbf {\bibinfo {volume}
  {87}},\ \bibinfo {pages} {115122} (\bibinfo {year} {2013})}\BibitemShut
  {NoStop}%
\bibitem [{\citenamefont {Van~Kampen}(2007)}]{VanKampen2007}%
  \BibitemOpen
  \bibfield  {author} {\bibinfo {author} {\bibfnamefont {N.~G.}\ \bibnamefont
  {Van~Kampen}},\ }\href@noop {} {\emph {\bibinfo {title} {Stochastic Processes
  in Physics and Chemistry}}},\ \bibinfo {edition} {3rd}\ ed.\ (\bibinfo
  {publisher} {North-Holland},\ \bibinfo {address} {Amsterdam},\ \bibinfo
  {year} {2007})\BibitemShut {NoStop}%
\bibitem [{\citenamefont {Gillespie}(1991)}]{Gillespie1991}%
  \BibitemOpen
  \bibfield  {author} {\bibinfo {author} {\bibfnamefont {D.~T.}\ \bibnamefont
  {Gillespie}},\ }\href@noop {} {\emph {\bibinfo {title} {Markov Processes: An
  Introduction for Physical Scientists}}},\ \bibinfo {edition} {1st}\ ed.\
  (\bibinfo  {publisher} {Academic Press},\ \bibinfo {address} {New York},\
  \bibinfo {year} {1991})\BibitemShut {NoStop}%
\bibitem [{\citenamefont {Gillespie}(1996{\natexlab{a}})}]{Gillespie1995}%
  \BibitemOpen
  \bibfield  {author} {\bibinfo {author} {\bibfnamefont {D.~T.}\ \bibnamefont
  {Gillespie}},\ }\bibfield  {title} {\enquote {\bibinfo {title} {Exact
  numerical simulation of the \text{O}rnstein-\text{U}hlenbeck process and its
  integral},}\ }\href {\doibase 10.1103/PhysRevE.54.2084} {\bibfield  {journal}
  {\bibinfo  {journal} {Phys. Rev. E}\ }\textbf {\bibinfo {volume} {54}},\
  \bibinfo {pages} {2084} (\bibinfo {year} {1996}{\natexlab{a}})}\BibitemShut
  {NoStop}%
\bibitem [{\citenamefont {Gillespie}(1996{\natexlab{b}})}]{Gillespie1996}%
  \BibitemOpen
  \bibfield  {author} {\bibinfo {author} {\bibfnamefont {D.~T.}\ \bibnamefont
  {Gillespie}},\ }\bibfield  {title} {\enquote {\bibinfo {title} {The
  mathematics of \text{B}rownian motion and \text{J}ohnson noise},}\ }\href
  {\doibase 10.1119/1.18210} {\bibfield  {journal} {\bibinfo  {journal} {Am. J.
  Phys.}\ }\textbf {\bibinfo {volume} {64}},\ \bibinfo {pages} {225} (\bibinfo
  {year} {1996}{\natexlab{b}})}\BibitemShut {NoStop}%
\end{thebibliography}

\begin{thebibliography}{15}%
\makeatletter
\providecommand \@ifxundefined [1]{%
\@ifx{#1\undefined}
}%
\providecommand \@ifnum [1]{%
\ifnum #1\expandafter \@firstoftwo
\else \expandafter \@secondoftwo
\fi
}%
\providecommand \@ifx [1]{%
\ifx #1\expandafter \@firstoftwo
\else \expandafter \@secondoftwo
\fi
}%
\providecommand \natexlab [1]{#1}%
\providecommand \enquote  [1]{``#1''}%
\providecommand \bibnamefont  [1]{#1}%
\providecommand \bibfnamefont [1]{#1}%
\providecommand \citenamefont [1]{#1}%
\providecommand \href@noop [0]{\@secondoftwo}%
\providecommand \href [0]{\begingroup \@sanitize@url \@href}%
\providecommand \@href[1]{\@@startlink{#1}\@@href}%
\providecommand \@@href[1]{\endgroup#1\@@endlink}%
\providecommand \@sanitize@url [0]{\catcode `\\12\catcode `\$12\catcode
`\&12\catcode `\#12\catcode `\^12\catcode `\_12\catcode `\%12\relax}%
\providecommand \@@startlink[1]{}%
\providecommand \@@endlink[0]{}%
\providecommand \url  [0]{\begingroup\@sanitize@url \@url }%
\providecommand \@url [1]{\endgroup\@href {#1}{\urlprefix }}%
\providecommand \urlprefix  [0]{URL }%
\providecommand \Eprint [0]{\href }%
\providecommand \doibase [0]{http://dx.doi.org/}%
\providecommand \selectlanguage [0]{\@gobble}%
\providecommand \bibinfo  [0]{\@secondoftwo}%
\providecommand \bibfield  [0]{\@secondoftwo}%
\providecommand \translation [1]{[#1]}%
\providecommand \BibitemOpen [0]{}%
\providecommand \bibitemStop [0]{}%
\providecommand \bibitemNoStop [0]{.\EOS\space}%
\providecommand \EOS [0]{\spacefactor3000\relax}%
\providecommand \BibitemShut  [1]{\csname bibitem#1\endcsname}%
\let\auto@bib@innerbib\@empty
\bibitem [{\citenamefont {Kolkowitz}\ \emph {et~al.}(2012)\citenamefont
{Kolkowitz}, \citenamefont {Unterreithmeier}, \citenamefont {Bennett},\ and\
\citenamefont {Lukin}}]{sKolkowitz2012}%
\BibitemOpen
\bibfield  {author} {\bibinfo {author} {\bibfnamefont {S.}~\bibnamefont
{Kolkowitz}}, \bibinfo {author} {\bibfnamefont {Q.~P.}\ \bibnamefont
{Unterreithmeier}}, \bibinfo {author} {\bibfnamefont {S.~D.}\ \bibnamefont
{Bennett}}, \ and\ \bibinfo {author} {\bibfnamefont {M.~D.}\ \bibnamefont
{Lukin}},\ }\bibfield  {title} {\enquote {\bibinfo {title} {Sensing distant
nuclear spins with a single electron spin},}\ }\href {\doibase
10.1103/PhysRevLett.109.137601} {\bibfield  {journal} {\bibinfo  {journal}
{Phys. Rev. Lett.}\ }\textbf {\bibinfo {volume} {109}},\ \bibinfo {pages}
{137601} (\bibinfo {year} {2012})}\BibitemShut {NoStop}%
\bibitem [{\citenamefont {Taminiau}\ \emph {et~al.}(2012)\citenamefont
{Taminiau}, \citenamefont {Wagenaar}, \citenamefont {van~der Sar},
\citenamefont {Jelezko}, \citenamefont {Dobrovitski},\ and\ \citenamefont
{Hanson}}]{sTaminiau2012}%
\BibitemOpen
\bibfield  {author} {\bibinfo {author} {\bibfnamefont {T.~H.}\ \bibnamefont
{Taminiau}}, \bibinfo {author} {\bibfnamefont {J.~J.~T.}\ \bibnamefont
{Wagenaar}}, \bibinfo {author} {\bibfnamefont {T.}~\bibnamefont {van~der
Sar}}, \bibinfo {author} {\bibfnamefont {F.}~\bibnamefont {Jelezko}},
\bibinfo {author} {\bibfnamefont {V.~V.}\ \bibnamefont {Dobrovitski}}, \ and\
\bibinfo {author} {\bibfnamefont {R.}~\bibnamefont {Hanson}},\ }\bibfield
{title} {\enquote {\bibinfo {title} {Detection and control of individual
nuclear spins using a weakly coupled electron spin},}\ }\href {\doibase
10.1103/PhysRevLett.109.137602} {\bibfield  {journal} {\bibinfo  {journal}
{Phys. Rev. Lett.}\ }\textbf {\bibinfo {volume} {109}},\ \bibinfo {pages}
{137602} (\bibinfo {year} {2012})}\BibitemShut {NoStop}%
\bibitem [{\citenamefont {Grinolds}\ \emph {et~al.}(2014)\citenamefont
{Grinolds}, \citenamefont {Warner}, \citenamefont {De~Greve}, \citenamefont
{Dovzhenko}, \citenamefont {Thiel}, \citenamefont {Walsworth}, \citenamefont
{Hong}, \citenamefont {Maletinsky},\ and\ \citenamefont
{Yacoby}}]{sGrinolds2014}%
\BibitemOpen
\bibfield  {author} {\bibinfo {author} {\bibfnamefont {M.~S.}\ \bibnamefont
{Grinolds}}, \bibinfo {author} {\bibfnamefont {M.}~\bibnamefont {Warner}},
\bibinfo {author} {\bibfnamefont {K.}~\bibnamefont {De~Greve}}, \bibinfo
{author} {\bibfnamefont {Y.}~\bibnamefont {Dovzhenko}}, \bibinfo {author}
{\bibfnamefont {L.}~\bibnamefont {Thiel}}, \bibinfo {author} {\bibfnamefont
{R.~L.}\ \bibnamefont {Walsworth}}, \bibinfo {author} {\bibfnamefont
{S.}~\bibnamefont {Hong}}, \bibinfo {author} {\bibfnamefont {P.}~\bibnamefont
{Maletinsky}}, \ and\ \bibinfo {author} {\bibfnamefont {A.}~\bibnamefont
{Yacoby}},\ }\bibfield  {title} {\enquote {\bibinfo {title} {Subnanometre
resolution in three-dimensional magnetic resonance imaging of individual dark
spins},}\ }\href {\doibase 10.1038/NNANO.2014.30} {\bibfield  {journal}
{\bibinfo  {journal} {Nat. Nanotechnol.}\ }\textbf {\bibinfo {volume} {9}},\
\bibinfo {pages} {279} (\bibinfo {year} {2014})}\BibitemShut {NoStop}%
\bibitem [{\citenamefont {Zopes}\ \emph {et~al.}(2018)\citenamefont {Zopes},
\citenamefont {Cujia}, \citenamefont {Sasaki}, \citenamefont {Boss},
\citenamefont {Itoh},\ and\ \citenamefont {Degen}}]{sZopes2018}%
\BibitemOpen
\bibfield  {author} {\bibinfo {author} {\bibfnamefont {J.}~\bibnamefont
{Zopes}}, \bibinfo {author} {\bibfnamefont {K.~S.}\ \bibnamefont {Cujia}},
\bibinfo {author} {\bibfnamefont {K.}~\bibnamefont {Sasaki}}, \bibinfo
{author} {\bibfnamefont {J.~M.}\ \bibnamefont {Boss}}, \bibinfo {author}
{\bibfnamefont {K.~M.}\ \bibnamefont {Itoh}}, \ and\ \bibinfo {author}
{\bibfnamefont {C.~L.}\ \bibnamefont {Degen}},\ }\bibfield  {title} {\enquote
{\bibinfo {title} {Three-dimensional localization spectroscopy of individual
nuclear spins with sub-angstrom resolution},}\ }\href {\doibase
10.1038/s41467-018-07121-0} {\bibfield  {journal} {\bibinfo  {journal} {Nat.
Commun.}\ }\textbf {\bibinfo {volume} {9}},\ \bibinfo {pages} {4678}
(\bibinfo {year} {2018})}\BibitemShut {NoStop}%
\bibitem [{\citenamefont {Dobrovitski}\ \emph {et~al.}(2008)\citenamefont
{Dobrovitski}, \citenamefont {Feiguin}, \citenamefont {Awschalom},\ and\
\citenamefont {Hanson}}]{sDobrovitski2008}%
\BibitemOpen
\bibfield  {author} {\bibinfo {author} {\bibfnamefont {V.~V.}\ \bibnamefont
{Dobrovitski}}, \bibinfo {author} {\bibfnamefont {A.~E.}\ \bibnamefont
{Feiguin}}, \bibinfo {author} {\bibfnamefont {D.~D.}\ \bibnamefont
{Awschalom}}, \ and\ \bibinfo {author} {\bibfnamefont {R.}~\bibnamefont
{Hanson}},\ }\bibfield  {title} {\enquote {\bibinfo {title} {Decoherence
dynamics of a single spin versus spin ensemble},}\ }\href {\doibase
10.1103/PhysRevB.77.245212} {\bibfield  {journal} {\bibinfo  {journal} {Phys.
Rev. B}\ }\textbf {\bibinfo {volume} {77}},\ \bibinfo {pages} {245212}
(\bibinfo {year} {2008})}\BibitemShut {NoStop}%
\bibitem [{\citenamefont {Dobrovitski}\ \emph {et~al.}(2009)\citenamefont
{Dobrovitski}, \citenamefont {Feiguin}, \citenamefont {Hanson},\ and\
\citenamefont {Awschalom}}]{sDobrovitski2009}%
\BibitemOpen
\bibfield  {author} {\bibinfo {author} {\bibfnamefont {V.~V.}\ \bibnamefont
{Dobrovitski}}, \bibinfo {author} {\bibfnamefont {A.~E.}\ \bibnamefont
{Feiguin}}, \bibinfo {author} {\bibfnamefont {R.}~\bibnamefont {Hanson}}, \
and\ \bibinfo {author} {\bibfnamefont {D.~D.}\ \bibnamefont {Awschalom}},\
}\bibfield  {title} {\enquote {\bibinfo {title} {Decay of \text{R}abi
oscillations by dipolar-coupled dynamical spin environments},}\ }\href
{\doibase 10.1103/PhysRevLett.102.237601} {\bibfield  {journal} {\bibinfo
{journal} {Phys. Rev. Lett.}\ }\textbf {\bibinfo {volume} {102}},\ \bibinfo
{pages} {237601} (\bibinfo {year} {2009})}\BibitemShut {NoStop}%
\bibitem [{\citenamefont {de~Lange}\ \emph {et~al.}(2010)\citenamefont
{de~Lange}, \citenamefont {Wang}, \citenamefont {Rist{\`e}}, \citenamefont
{Dobrovitski},\ and\ \citenamefont {Hanson}}]{sde-Lange2010}%
\BibitemOpen
\bibfield  {author} {\bibinfo {author} {\bibfnamefont {G.}~\bibnamefont
{de~Lange}}, \bibinfo {author} {\bibfnamefont {Z.~H.}\ \bibnamefont {Wang}},
\bibinfo {author} {\bibfnamefont {D.}~\bibnamefont {Rist{\`e}}}, \bibinfo
{author} {\bibfnamefont {V.~V.}\ \bibnamefont {Dobrovitski}}, \ and\ \bibinfo
{author} {\bibfnamefont {R.}~\bibnamefont {Hanson}},\ }\bibfield  {title}
{\enquote {\bibinfo {title} {Universal dynamical decoupling of a single
solid-state spin from a spin bath},}\ }\href {\doibase
10.1126/science.1192739} {\bibfield  {journal} {\bibinfo  {journal}
{Science}\ }\textbf {\bibinfo {volume} {330}},\ \bibinfo {pages} {60}
(\bibinfo {year} {2010})}\BibitemShut {NoStop}%
\bibitem [{\citenamefont {Maze}\ \emph {et~al.}(2012)\citenamefont {Maze},
\citenamefont {Dréau}, \citenamefont {Waselowski}, \citenamefont {Duarte},
\citenamefont {Roch},\ and\ \citenamefont {Jacques}}]{sMaze2012}%
\BibitemOpen
\bibfield  {author} {\bibinfo {author} {\bibfnamefont {J.~R.}\ \bibnamefont
{Maze}}, \bibinfo {author} {\bibfnamefont {A.}~\bibnamefont {Dréau}},
\bibinfo {author} {\bibfnamefont {V.}~\bibnamefont {Waselowski}}, \bibinfo
{author} {\bibfnamefont {H.}~\bibnamefont {Duarte}}, \bibinfo {author}
{\bibfnamefont {J.-F.}\ \bibnamefont {Roch}}, \ and\ \bibinfo {author}
{\bibfnamefont {V}~\bibnamefont {Jacques}},\ }\bibfield  {title} {\enquote
{\bibinfo {title} {Free induction decay of single spins in diamond},}\ }\href
{\doibase 10.1088/1367-2630/14/10/103041} {\bibfield  {journal} {\bibinfo
{journal} {New J. Phys.}\ }\textbf {\bibinfo {volume} {14}},\ \bibinfo
{pages} {103041} (\bibinfo {year} {2012})}\BibitemShut {NoStop}%
\bibitem [{\citenamefont {Wang}\ and\ \citenamefont
{Takahashi}(2013)}]{sWang2013}%
\BibitemOpen
\bibfield  {author} {\bibinfo {author} {\bibfnamefont {Z.-H.}\ \bibnamefont
{Wang}}\ and\ \bibinfo {author} {\bibfnamefont {S.}~\bibnamefont
{Takahashi}},\ }\bibfield  {title} {\enquote {\bibinfo {title} {Spin
decoherence and electron spin bath noise of a nitrogen-vacancy center in
diamond},}\ }\href {\doibase 10.1103/PhysRevB.87.115122} {\bibfield
{journal} {\bibinfo  {journal} {Phys. Rev. B}\ }\textbf {\bibinfo {volume}
{87}},\ \bibinfo {pages} {115122} (\bibinfo {year} {2013})}\BibitemShut
{NoStop}%
\bibitem [{\citenamefont {Van~Kampen}(2007)}]{sVanKampen2007}%
\BibitemOpen
\bibfield  {author} {\bibinfo {author} {\bibfnamefont {N.~G.}\ \bibnamefont
{Van~Kampen}},\ }\href@noop {} {\emph {\bibinfo {title} {Stochastic Processes
in Physics and Chemistry}}},\ \bibinfo {edition} {3rd}\ ed.\ (\bibinfo
{publisher} {North-Holland},\ \bibinfo {address} {Amsterdam},\ \bibinfo
{year} {2007})\BibitemShut {NoStop}%
\bibitem [{\citenamefont {Gillespie}(1991)}]{sGillespie1991}%
\BibitemOpen
\bibfield  {author} {\bibinfo {author} {\bibfnamefont {D.~T.}\ \bibnamefont
{Gillespie}},\ }\href@noop {} {\emph {\bibinfo {title} {Markov Processes: An
Introduction for Physical Scientists}}},\ \bibinfo {edition} {1st}\ ed.\
(\bibinfo  {publisher} {Academic Press},\ \bibinfo {address} {New York},\
\bibinfo {year} {1991})\BibitemShut {NoStop}%
\bibitem [{\citenamefont {Gillespie}(1996{\natexlab{a}})}]{sGillespie1995}%
\BibitemOpen
\bibfield  {author} {\bibinfo {author} {\bibfnamefont {D.~T.}\ \bibnamefont
{Gillespie}},\ }\bibfield  {title} {\enquote {\bibinfo {title} {Exact
numerical simulation of the \text{O}rnstein-\text{U}hlenbeck process and its
integral},}\ }\href {\doibase 10.1103/PhysRevE.54.2084} {\bibfield  {journal}
{\bibinfo  {journal} {Phys. Rev. E}\ }\textbf {\bibinfo {volume} {54}},\
\bibinfo {pages} {2084} (\bibinfo {year} {1996}{\natexlab{a}})}\BibitemShut
{NoStop}%
\bibitem [{\citenamefont {Gillespie}(1996{\natexlab{b}})}]{sGillespie1996}%
\BibitemOpen
\bibfield  {author} {\bibinfo {author} {\bibfnamefont {D.~T.}\ \bibnamefont
{Gillespie}},\ }\bibfield  {title} {\enquote {\bibinfo {title} {The
mathematics of \text{B}rownian motion and \text{J}ohnson noise},}\ }\href
{\doibase 10.1119/1.18210} {\bibfield  {journal} {\bibinfo  {journal} {Am. J.
Phys.}\ }\textbf {\bibinfo {volume} {64}},\ \bibinfo {pages} {225--240}
(\bibinfo {year} {1996}{\natexlab{b}})}\BibitemShut {NoStop}%
\bibitem [{\citenamefont {Naydenov}\ \emph {et~al.}(2011)\citenamefont
{Naydenov}, \citenamefont {Dolde}, \citenamefont {Hall}, \citenamefont
{Shin}, \citenamefont {Fedder}, \citenamefont {Hollenberg}, \citenamefont
{Jelezko},\ and\ \citenamefont {Wrachtrup}}]{sNaydenov2011}%
\BibitemOpen
\bibfield  {author} {\bibinfo {author} {\bibfnamefont {B.}~\bibnamefont
{Naydenov}}, \bibinfo {author} {\bibfnamefont {F.}~\bibnamefont {Dolde}},
\bibinfo {author} {\bibfnamefont {L.~T.}\ \bibnamefont {Hall}}, \bibinfo
{author} {\bibfnamefont {C.}~\bibnamefont {Shin}}, \bibinfo {author}
{\bibfnamefont {H.}~\bibnamefont {Fedder}}, \bibinfo {author} {\bibfnamefont
{L.~C.~L.}\ \bibnamefont {Hollenberg}}, \bibinfo {author} {\bibfnamefont
{F.}~\bibnamefont {Jelezko}}, \ and\ \bibinfo {author} {\bibfnamefont
{J.}~\bibnamefont {Wrachtrup}},\ }\bibfield  {title} {\enquote {\bibinfo
{title} {Dynamical decoupling of a single-electron spin at room
temperature},}\ }\href {\doibase 10.1103/PhysRevB.83.081201} {\bibfield
{journal} {\bibinfo  {journal} {Phys. Rev. B}\ }\textbf {\bibinfo {volume}
{83}},\ \bibinfo {pages} {081201} (\bibinfo {year} {2011})}\BibitemShut
{NoStop}%
\bibitem [{\citenamefont {Cahill}\ and\ \citenamefont
{Glauber}(1969)}]{sCahill1969a}%
\BibitemOpen
\bibfield  {author} {\bibinfo {author} {\bibfnamefont {K.~E.}\ \bibnamefont
{Cahill}}\ and\ \bibinfo {author} {\bibfnamefont {R.~J.}\ \bibnamefont
{Glauber}},\ }\bibfield  {title} {\enquote {\bibinfo {title} {Ordered
expansions in boson amplitude operators},}\ }\href {\doibase
10.1103/PhysRev.177.1857} {\bibfield  {journal} {\bibinfo  {journal} {Phys.
Rev.}\ }\textbf {\bibinfo {volume} {177}},\ \bibinfo {pages} {1857} (\bibinfo
{year} {1969})}\BibitemShut {NoStop}%
\end{thebibliography}
\end{document}